\documentclass[final,3p,times,twocolumn]{elsarticle}
\usepackage{hyperref}

\usepackage{amsmath,amssymb}
\usepackage[version=4]{mhchem}  
\usepackage{siunitx}

\usepackage{ulem}

\newcommand{\overbar}[1]{\mkern 1.5mu\overline{\mkern-1.5mu#1\mkern-1.5mu}\mkern 1.5mu}

\usepackage{graphicx}
\usepackage{color}

\usepackage{natbib}
\setcitestyle{square,numbers}

\usepackage[inline]{enumitem}
\usepackage{caption}
\usepackage{subcaption}

\usepackage{booktabs}
\usepackage{multirow}

\usepackage{listings}
\usepackage{color}
\usepackage{comment}

\renewcommand{\vec}[1]{\mathbf{#1}}

\usepackage{todonotes}

\begin{document}

\begin{frontmatter}

\title{Physics-based battery model parametrisation from impedance data}

\author[inst1,inst4]{Noël Hallemans\corref{cor1}}
\cortext[cor1]{Corresponding author} 
\ead{noel.hallemans@eng.ox.ac.uk}
\author[inst1,inst4]{Nicola E. Courtier}
\author[inst2,inst4]{Colin P. Please}
\author[inst1]{Brady Planden}
\author[inst2]{Rishit Dhoot}
\author[inst3]{Robert Timms}
\author[inst2,inst4]{S. Jon Chapman}
\author[inst1,inst4]{David Howey}
\author[inst1,inst4]{Stephen R. Duncan}
\affiliation[inst1]{
            organization={Department of Engineering Science, University of Oxford},
            city={Oxford},
            postcode={OX1 3PJ},
            country={UK}
            }
\affiliation[inst2]{
            organization={Mathematical Institute, University of Oxford},
            addressline={Andrew Wiles Building, Woodstock Road},
            city={Oxford},
            postcode={OX2 6GG},
            country={UK}
            }
\affiliation[inst3]{
            organization={Ionworks Technologies Inc},
           addressline={5831 Forward Ave \#1276 },
            city={Pittsburgh, PA},
            postcode={15217},
            country={USA}
          }
\affiliation[inst4]{
            organization={The Faraday Institution},
            addressline={Quad One, Becquerel Avenue, Harwell Campus},
            city={Didcot},
            postcode={OX11 0RA},
            country={UK}
            }



\begin{abstract}
Non-invasive parametrisation of physics-based battery models can be performed by fitting the model to electrochemical impedance spectroscopy (EIS) data containing features related to the different physical processes. However, this requires an impedance model to be derived, which may be complex to obtain analytically. We have developed the open-source software PyBaMM-EIS that provides a fast method to compute the impedance of any PyBaMM model at any operating point using automatic differentiation. Using PyBaMM-EIS, we investigate the impedance of the single particle model, single particle model with electrolyte (SPMe), and Doyle-Fuller-Newman model, and identify the SPMe as a parsimonious option that shows the typical features of measured lithium-ion cell impedance data. We provide a grouped-parameter SPMe and analyse the features in the impedance related to each parameter. Using the open-source software PyBOP, we estimate 18 grouped parameters both from simulated impedance data and from measured impedance data from a LG M50LT lithium-ion battery. The parameters that directly affect the response of the SPMe can be accurately determined and assigned to the correct electrode. Crucially, parameter fitting must be done simultaneously to measurements across a wide range of states-of-charge. Overall, this work presents a practical way to find the parameters of physics-based models.

\end{abstract}
\end{frontmatter}

\setlength{\abovecaptionskip}{5pt} 
Characterisation techniques are essential for understanding physical processes in batteries and monitoring their state. In many applications, such as health estimation in battery management systems (BMS) and recycling facilities, characterisation should be non-invasive and based on current and voltage data. Characterising a battery is often performed by parametrising a model (a relation between terminal voltage and applied current) from measured data, which can then be used for physical interpretation and simulation of the battery. 

The choice of the model may be challenging. Often ad hoc equivalent circuit models \cite{plett2015battery} are used to fit measured data \cite{hu2012comparative,lagnoni2024critical}, but these have the disadvantage that the parameters are not always physically meaningful and simulation may be inaccurate. Instead, we choose to parametrise \textit{physics-based} models \cite{plett2023battery,tian2025physics}. The standard physics-based modelling approach for lithium-ion batteries is the Doyle-Fuller-Newman (DFN) framework \cite{doyle1993modeling,doyle1996comparison}, based on porous-electrode theory, consisting of a set of coupled nonlinear partial differential equations that are typically too expensive to simulate in a BMS. There also exist reduced-order models, derived from the DFN model, such as the single-particle model (SPM) \cite{guo2010single}, single-particle model with electrolyte (SPMe) \cite{moura2016battery,marquis2019,planella2021systematic}, or multi-particle model \cite{kirk2022modeling}, all of which may be more practical for BMS. 

Model parameters are only identifiable from measured datasets that are sufficiently informative, containing information at the relevant time-scales of the physical processes in the model---which may be orders of magnitude apart. One approach uses constant-current or pulse data during cycling to fit the model to the time-domain voltage response, which we refer to as the ``time-domain voltage method'' \cite{lu2021nondestructive}. However, this is often not sufficiently informative to fully parametrise a physics-based model as the excitation may not excite all time-scales and requires large datasets.  Electrochemical impedance spectroscopy (EIS) is therefore often used as a complementary source of data, providing compressed information over a wide range of time-scales (or, equivalently, frequencies) \cite{plett2023battery,iurilli2021use,hileman2024estimating,zhu2024study,wimarshana2023multi}. Conventional EIS measurements \cite{EISbook,wang2021electrochemical,vivier2022impedance} consist of taking an equilibrated battery at a fixed operating point (state-of-charge (SOC) and temperature) and applying a small sinusoidal current (or voltage) at set amplitude and frequency. The resulting voltage (or current) is measured and the Fourier component of this, at the same frequency as the excitation, is used to calculate the impedance. The process is then repeated at several frequencies. Measuring impedance over a wide frequency range allows physical processes occurring at different time-scales to be distinguished, and including impedance data for model parametrisation has been shown to improve the identifiability of parameters \cite{lu2022nondestructive}. The frequencies are typically spaced logarithmically to give appropriate weights to the different physical processes in the impedance data.

To parametrise a physics-based model from impedance data, we need to be able to compute the impedance response of the model. There are two basic approaches, one that solves the problem in the time-domain and the other in the frequency-domain.

In the first method, the impedance can be computed directly in the time-domain (the ``brute-force" approach) by simulating the conventional EIS measurement procedure described above on a model. The advantages of this are that it is easy to implement, the simulation is true to the practical method, and additional information can be extracted, if required. For example, the higher harmonics can be examined for use in nonlinear EIS (NLEIS) \cite{murbach2018nonlinear,kirk2023nonlinear,ji2024second}. The main drawback of the brute-force method is relatively high computational cost.

In the second approach, the model is first linearised about an operating point before being transformed into the frequency-domain, which can sometimes be performed analytically from the model equations. Early examples of this approach include the impedance calculations in Lasia et al.\ \cite{lasia1995impedance} and Meyers et al.\ \cite{meyers2000}, and the impedance mode of the classical pseudo two-dimensional Dualfoil code \cite{albertus2007}. More recently, Song and Bazant analytically calculated the diffusion impedance of battery electrodes for different particle geometries \cite{song2012effects}, Bizeray et al.\ calculated the impedance of the full SPM~\cite{bizeray2018identifiability}, and Plett and Trimboli studied the impedance of the DFN \cite{plett2023battery}.
However, practically extending this to more complicated models is challenging, so numerical approaches are adopted on the discretised model in space (usually using finite differences or finite volumes). One method, referred to as numerical differentiation, uses a finite difference approximation for the derivatives arising from the linearisation, such as in Zhu et al.\ \cite{zhu2024extracting}. Alternatively, here we create exact expressions for the derivatives using \textit{automatic differentiation}, leveraging the open-source battery simulation software \texttt{PyBaMM} \cite{sulzer2021python}, which stores models as analytical equations. Automatic differentiation---implemented in \texttt{JAX} \cite{jax2018github}, for instance---computes derivatives by applying the chain rule to the algebraic operations and analytical functions within the model, and has been used previously on specific models \cite{zic_2020,huang2024minn}. Using these expressions, the impedance of the model can be computed numerically at \textit{any} operating point at a set of frequencies, and we have developed an open-source Python package, \texttt{PyBaMM-EIS}, that enables simulation of EIS with \textit{any} \texttt{PyBaMM} model. This ``numerical frequency-domain'' impedance method has a much lower computational cost than the brute-force method and is therefore suitable for model parametrisation.

Using \texttt{PyBaMM-EIS}, we can easily compare the impedance of many commonly used models (SPM, SPMe, and DFN). We found that including electrolyte dynamics adds a ``diffusion bump'' impedance feature, often present in measured battery data. We study the impedance of the SPMe in more detail and provide a grouped-parameter model to explain how its impedance changes with SOC, and to show the effect of each of the grouped parameters on the impedance features. Using the open-source software \texttt{PyBOP} \cite{PyBOP2024}, we study the parametrisation of the SPMe from simulated impedance and voltage data. Finally, we parametrise the SPMe from measured data of a LG~M50LT battery, however, obtaining good fits at low frequencies turns out to be difficult, showing that extensions of the SPMe are required to fit measured impedance data well. In spite of these shortcomings, as we shall show, the parametrised SPMe for the LG M50LT performs well for a drive cycle validation.

\section{Parametrising battery models}
\begin{figure}
    \centering
    \includegraphics[width=0.45\textwidth]{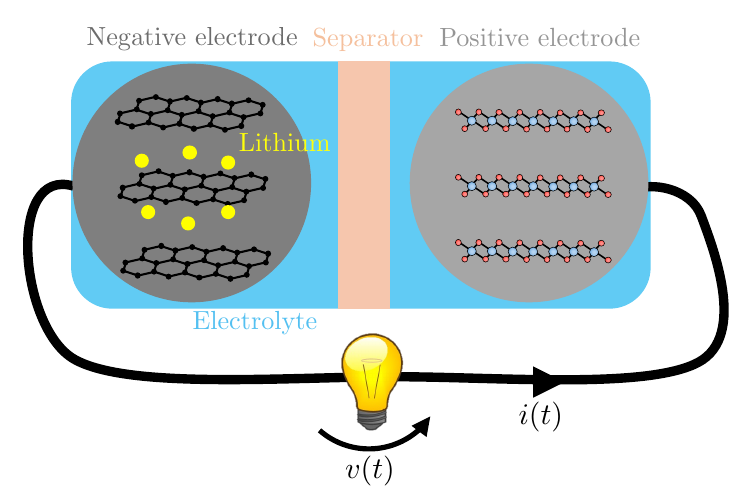}
    \caption{Illustration of a physics-based battery model with single particles and electrolyte.}
    \label{Fig:illustrationModel}
\end{figure}
A battery model (Fig.~\ref{Fig:illustrationModel}) allows us to simulate the terminal voltage $v(t)$ as an operator acting on the applied current $i(t)$ and depending on parameters $\theta$,
\begin{align}
    v(t)=F_\theta\{i(t)\},
\end{align}
with the variable $t$ representing time and $F_\theta$ acting on $i(\tau)$ with $\tau\in (-\infty,t]$. In this work, we look at physics-based models, such as the SPM, SPMe, or DFN. These relate the terminal voltage to the applied current based on processes occurring within the battery such as diffusion and charge transfer. Physics-based models depend on a set of parameters $\theta$ (Table~\ref{tab:groupedParamsSPM} for the SPMe) and consist of systems of partial differential equations (in space and time), ordinary differential equations, differential algebraic equations (DAEs), and nonlinear operators, as set out in Appendix~A for the SPMe. Note that the open circuit potentials (OCPs) of each electrode are also required, and account for a major part of the model. Non-invasive model parametrisation for a specific battery consists of estimating the parameters $\theta$ from current and voltage data from that battery.

\subsection{Time-domain voltage method}
\begin{figure}
    \centering
    \includegraphics[width=0.5\textwidth]{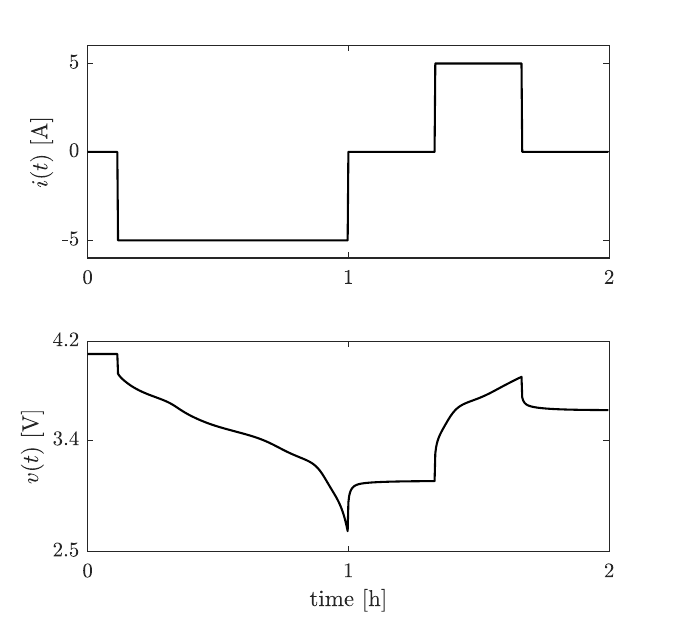}
    \caption{Time-domain simulation of the SPMe in Appendix~C. Top: applied current $i(t)$. Bottom: simulated voltage response $v(t)$.} 
    \label{Fig:timeDomainSimulation}
\end{figure}

Before we look at EIS data, we briefly consider the time-domain method where we parametrise a model by specifying the current and minimising the difference between measured and simulated voltage. An example of such data is in Fig.~\ref{Fig:timeDomainSimulation}, and one might choose optimal parameter values $\hat{\theta}$ to satisfy 
\begin{align}
    \hat \theta = \arg \min_\theta \sum_{n=0}^{N-1}\big(v(t_n)-v(t_n,i(t),\theta,\mathrm{SOC}_0)\big)^2,
    \label{eq:timeDomainOptimisation}
\end{align}
with $v(t_n)$ the measured data at discrete times $t_n$ and $v(t_n,i(t),\theta,\mathrm{SOC}_0)$ the voltage simulated by the model with initial condition $\mathrm{SOC}_0$ (SOC at time $t_0$). 

In general, when fitting model parameters to data, an alternative to the frequentist approach is the Bayesian approach, which has been used to explore the identifiability of the SPMe \cite{aitio2020bayesian}. Both of these approaches can be performed with the open source Python package \texttt{PyBOP} \cite{PyBOP2024}, which fits \texttt{PyBaMM} models to data. 

Physical processes in batteries occur at time-scales that may be several orders of magnitude apart. To ensure this is accounted for when parametrising a physics-based model, the time-domain data $v(t_n)$ should therefore cover these time-scales and $i(t)$ should excite the time-scales of interest sufficiently---which is not the case for typical charge and discharge data as in Fig.~\ref{Fig:timeDomainSimulation}. Drive cycles provide a more persistent excitation, but long records sampled at high frequency (large $N$) are required to cover the time-scales. EIS data, instead, provides an efficient data representation of the battery's dynamics over a wide range of time-scales, however, only valid for small perturbations of the voltage at operating points. In the next sections we detail how to measure impedance data and use it for model parametrisation.

\subsection{Measuring impedance data}
Impedance spectroscopy is a widely applied technique for parametrising models \cite{hu2023application}, not only in laboratories, but also in BMS. However, one has to be careful to measure \textit{valid} impedance data, that is, data satisfying the conditions of \textit{linearity} and \textit{stationarity} \cite{EISbook,hallemans2023electrochemical}.

A battery is in a stationary condition when it is at a fixed operating point $x_m$ (fixed SOC and temperature), typically after a relaxation period of 2 hours or longer (we use 4 hours). The condition of linearity is satisfied when the amplitudes of the sinusoidal current or voltage perturbations around the operating point are sufficiently small. A voltage deviation smaller than \SI{10}{mV} is often considered to indicate linearity, however this depends on several factors such as the frequency, temperature, and SOC. 

We choose the current as excitation,
\begin{align}
    i(t)=I_k \sin(\omega_k t),
    \label{eq:currentExcitation}
\end{align}
with $I_k$ the amplitude, $\omega_k=2\pi f_k$ the angular frequency, and $f_k$ the frequency. When linearity and stationarity are satisfied, the battery behaves as a linear time-invariant system around the operating point $x_m$. Hence, after transients have faded away, the voltage response to the sinusoidal current excitation \eqref{eq:currentExcitation} yields,
\begin{subequations}
\begin{align}
v(t)&=\text{OCV}_m+\mathcal{F}^{-1}\{Z_m(\omega)I(\omega)\}\\
    &=\text{OCV}_m+\underbrace{\vert Z_m(\omega_k)\vert I_k}_{V_k}\sin\big(\omega_k t + \underbrace{\angle Z_m(\omega_k)}_{\varphi_k}\big),
\end{align}
\end{subequations}
with OCV$_m$ the open circuit voltage at operating point $x_m$, $\mathcal{F}^{-1}\{\cdot\}$ the inverse Fourier transform, $Z_m(\omega)$ the impedance at operating point $x_m$, and $I(\omega)$ the Fourier transform of $i(t)$. The voltage response is thus also a sinusoidal signal, superimposed on the OCV, but with a different amplitude $V_k$ and a phase shift $\varphi_k$. The impedance at angular frequency $\omega_k$ can then be measured from the Fourier spectra of voltage and current as
\begin{align}
Z_m(\omega_k)=\frac{V(\omega_k)}{I(\omega_k)}=\frac{V_k}{I_k}e^{j\varphi_k},
\label{eq:impedanceDefinition}
\end{align}
where the complex impedance can be decomposed into its real part $Z_\mathrm{r}(\omega_k)$ and imaginary part $Z_\mathrm{j}(\omega_k)$, and $j$ is the imaginary unit ($j^2=-1$).

To obtain impedance data over a wide frequency range, this procedure is repeated at different angular frequencies $\omega_k$, which are typically logarithmically spaced. The conditions of linearity and stationarity should be checked for every frequency, which can be done by analysing the current and voltage data in the frequency-domain \cite{hallemans2023electrochemical}, looking at the Lissajous plots \cite{zabara2024utility}, or, when all impedance data is collected, by using the Kramers-Kronig relations \cite{urquidi1990applications} or a measurement model \cite{agarwal1995application,orazem2024measurement}. Note that for a battery, it is typically more difficult to obtain valid impedance data at low frequencies where nonlinear effects may be stronger \cite{hallemans2022operando}. 

\subsection{Parametrising a model from EIS data}
When measured over a wide frequency range, impedance data $Z_m(\omega_k)$ reveals information about physical processes occurring at different time-scales. Quantitative information about these processes can be obtained by fitting a model to the impedance data. 
\begin{figure}
    \centering
    \includegraphics[width=0.5\textwidth]{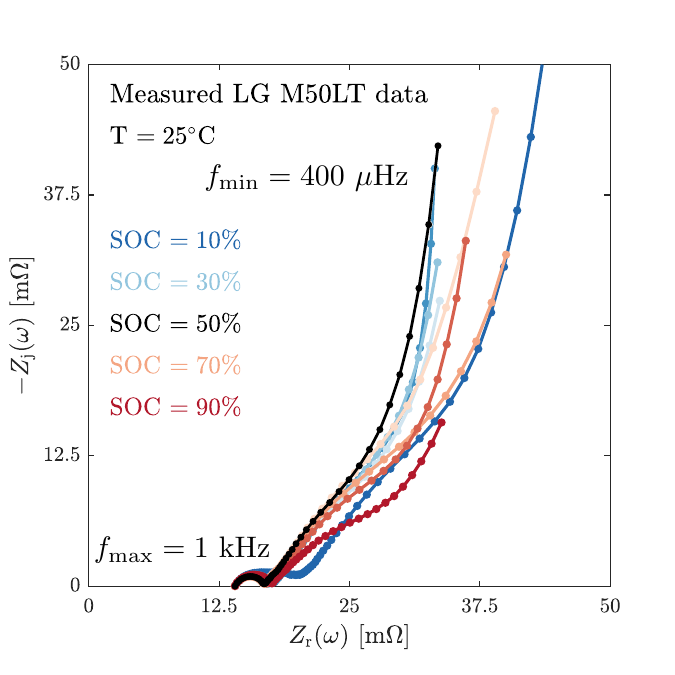}
    \caption{Nyquist plot of measured EIS data of the LG~M50LT at different SOCs and 25$^\circ$C.}
    \label{Fig:ZLGM50LTmeasured}
\end{figure}

To parametrise a physics-based model from measured impedance data we gather a dataset 
\begin{align}
   \mathcal{D}= \{\omega_k,Z_m(\omega_k)\}_{k=1,...,K}^{m=1,...,M},
   \label{eq:impedanceDataset}
\end{align}
where the index $k$ indicates excited frequencies and $m$ indicates operating points (e.g.\ different SOCs). An example of such a dataset with $M=9$ different SOC operating points and $K=65$ frequencies, logarithmically spaced between \SI{1}{kHz} and \SI{400}{\micro Hz}, is shown in Fig.~\ref{Fig:ZLGM50LTmeasured} for an LG~M50LT battery. Some of the model parameters $\theta$ can then be estimated by minimising a chosen cost function, for instance, the sum of squares of the difference between the impedance dataset and the model's impedance $Z_m(\omega,\theta)$,
\begin{align}
    \hat \theta = \arg \min_\theta \sum_{m=1}^M\sum_{k=1}^K\vert Z_m(\omega_k)-Z_m(\omega_k,\theta)\vert^2.
    \label{eq:costfunctionImpedance}
\end{align}
While it is common practice to estimate \textit{local} model parameters from impedance at single operating points (for instance for battery characterisation using local equivalent circuit models), we want to estimate \textit{global} model parameters to fit data at all the operating points. We do this by estimating the parameters simultaneously from impedance data at different SOCs.

For relatively simple models the impedance $Z_m(\omega,\theta)$ can be calculated analytically by linearising the model at the operating point $x_m$ and transforming it into the frequency-domain. However, for more complicated models, such as the SPMe or DFN, calculating an analytical expression of the impedance is prohibitively complicated. Therefore in this paper we propose a fast numerical frequency-domain method to obtain the model impedance.

\section{Computing numerical impedance with PyBaMM}
To simulate physics-based models, the spatial geometry (see Fig.~\ref{Fig:illustrationModel}) is discretised---for example using the finite volume method \cite{leveque2002}. It is important that sufficient discretisation points are chosen for accurate simulations; we take 100 radial discretisation points in the particles, 100 discretisation points over the thickness of each electrode, and 20 points over the thickness of the separator. Variables evaluated over this discretisation mesh (e.g.\ lithium concentration at specific points in the particles and electrolyte) are then converted into a state vector $\vec{x}(t)\in \mathbb{R}^{(N_\vec{x}+2)\times 1}$, with $N_\vec{x}$ the number of states in the \texttt{PyBaMM} model and the additional two states being the voltage and current. Spatial operators in the original model (such as gradients and divergences) become matrices in the discretised model. The model then becomes a system of DAEs, 
\begin{align}
    \vec{M}_\theta\big(\vec{x}(t)\big)\frac{\mathrm{d}\vec{x}(t)}{\mathrm{d}t} = \vec{F}_\theta\big(\vec{x}(t)\big) +
   \vec{B}i(t),
\label{eq:DAE}
\end{align}
with $\vec{M}_\theta$ the mass matrix, $\vec{F}_\theta$ a vector-valued nonlinear multivariate function, and $\vec{B}$ a zero column vector with a unit entry in its last element. It is this system of DAEs that is solved in battery modelling software such as \texttt{PyBaMM} \cite{sulzer2021python} to simulate the terminal voltage of a battery for a given applied current. 

To obtain the impedance of the model, the nonlinear system of DAEs \eqref{eq:DAE} can be linearised around an operating point $\vec{x}_m$ and transformed into the frequency-domain. The impedance at angular frequency $\omega_k$ can then be obtained as the scalar
\begin{align}
    Z_m(\omega_k,\theta)=\Big[(j\omega_k\vec{M}_{\theta,m}-\vec{J}_{\theta,m})^{-1}\vec{B}\Big]_{[N_\vec{x}+1]},
    \label{eq:NumericalImpedanceDef}
\end{align}
where $\vec{J}_{\theta,m}$ is the Jacobian of $\vec{F}_\theta$ at operating point $\vec{x}_m$, 
\begin{align}
    \vec{J}_{\theta,m}=\frac{\partial \vec{F}_\theta(\vec{x})}{\partial \vec{x}}\bigg\rvert_{\vec{x}=\vec{x}_m},
\end{align}
$\vec{M}_{\theta,m}=\vec{M}_\theta(\vec{x}_m)$ is the mass matrix evaluated at $\vec{x}_m$ (which is typically singular), and $N_\vec{x}+1$ indicates the index for the voltage in the vector. Note that the Jacobian being full rank makes the computation of the inverse in \eqref{eq:NumericalImpedanceDef} possible. The mathematical derivation for \eqref{eq:NumericalImpedanceDef} is given in Appendix~B.

We exploit \texttt{PyBaMM} to discretise the model and compute the Jacobian $\vec{J}_{\theta,m}$ using \texttt{JAX} automatic differentiation \cite{jax2018github}, with exact expressions for the derivatives. In addition, we have developed open-source software called \texttt{PyBaMM-EIS} that implements this numerical frequency-domain method, allowing efficient computation of the impedance of any battery model implemented in \texttt{PyBaMM}, at any operating point $\vec{x}_m$ during a simulation. Moreover, users can define their own model and compute an impedance from it.

\section{Validation of the numerical impedance}
To validate the proposed numerical frequency-domain impedance computation method it was compared to a brute-force simulation. The latter consists of simulating the voltage response of the model for a sinusoidal current input at different frequencies and computing the impedance \eqref{eq:impedanceDefinition} from the Fourier spectra of these. 

\begin{figure}
    \centering
    \includegraphics[width=0.5\textwidth]{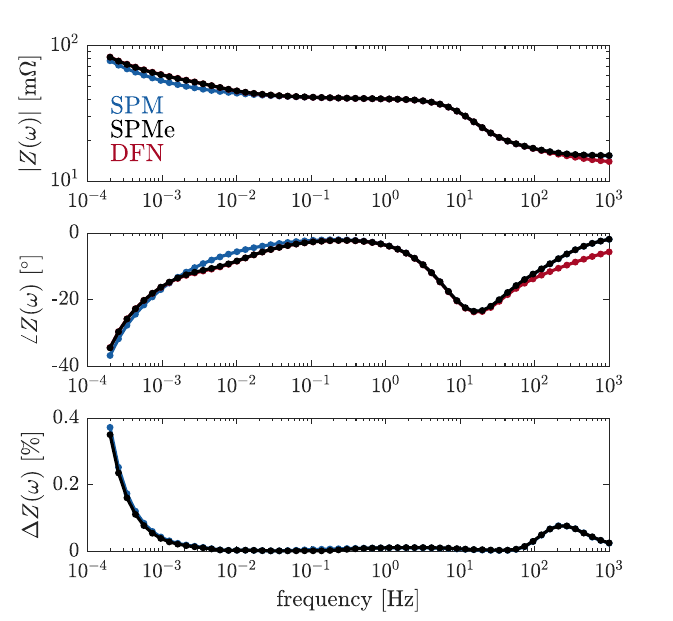}
    \caption{Bode plot of simulated impedance for the LG~M50 cell at 50\% SOC for the SPM, SPMe, and DFN computed using the numerical frequency-domain method, and relative difference between the frequency-domain and brute-force methods, $\Delta Z(\omega)\text{ [\%]} = 100\frac{\vert Z_\mathrm{bruteforce}(\omega) - Z_\mathrm{frequency}(\omega)\vert}{\vert Z_\mathrm{frequency}(\omega)\vert}$. The brute-force DFN computation time was prohibitively long, and, hence, we don't show the relative difference for this model.}
    \label{Fig:Zmethod_Bode}
\end{figure}

We consider the SPM, SPMe, and DFN models with double-layer capacitance implemented in \texttt{PyBaMM} (version \texttt{24.9.0}) with the parameter set of \texttt{chen2020} \cite{chen2020development} for the LG M50 cell. Fig.~\ref{Fig:Zmethod_Bode} shows a Bode plot of the EIS spectra of these models calculated using the frequency-domain approach and the relative difference compared to the brute-force approach. The models were discretised in space using the finite volume method resulting in a DAE system with the number of states $N_\vec{x}$ listed in Table~\ref{table:compareMethods}. The impedance response was computed for an operating point at 50\% SOC and \SI{25}{\celsius}, with 60 logarithmically spaced frequencies between \SI{200}{\micro \Hz} and \SI{1}{kHz}. For the frequency-domain numerical impedance, we used \texttt{PyBaMM-EIS}. For the brute-force approach, we chose a sinusoidal excitation with amplitude $I_k=$ \SI{100}{mA} and simulated the response for ten periods. To discard transient effects, only the five last periods were retained. To obtain accurate simulations over this wide frequency range, we set the absolute tolerance of the solver to $10^{-9}$. The impedance was computed from the ratio of voltage and current Fourier spectra \eqref{eq:impedanceDefinition}. 
\begin{table}
    \centering   
    \begin{tabular}{l l l l l}
    \hline
      \multirow{2}{*}{Model} & \multirow{2}{*}{$N_\vec{x}$} & \multicolumn{2}{c}{Computation time} \\
     &  & Brute-force & Freq. domain\\
     \hline
     SPM &  204 & 11.8 s & 21.3 ms\\   
     SPMe & 424 & 32.8 s & 415 ms\\   
     DFN & 20422 & -- & 925 ms\\     
     \hline   
     \end{tabular} 
    \caption{Number of states $N_\vec{x}$ and representative computation times for  impedance of different models and methods (averaged over ten runs). The brute-force DFN computation time was prohibitively long.}
    \label{table:compareMethods}         
 \end{table}

We observe that both methods have excellent agreement, with a relative error smaller than $0.4$\% for all models over the entire frequency range. Note that the frequency-domain approach computes the exact linearisation, while the brute-force approach is only exact when the amplitude of the sinusoidal excitation is small enough for linear response \cite{hallemans2023electrochemical}. Battery models are typically more nonlinear at low frequencies due to the nonlinear OCV, which explains the larger error at low frequencies. The computation times for obtaining the data in Fig.~\ref{Fig:Zmethod_Bode} are listed in Table~\ref{table:compareMethods} (note that the brute-force DFN simulation did not finish in a reasonable time due to the large number of states). We conclude that the numerical frequency-domain approach is faster and provides us with accurate impedance data.

\section{Comparing the impedance of different models}
We now compare the impedance response of several models using the \texttt{chen2020} LG~M50 cell parameter set. We needed 100 discretisation points in the particle to adequately resolve the impedance behaviour, especially in the higher frequency part of the diffusion region, and doing so made our results slightly differ from those of \cite{zhu2024extracting}. 

\begin{figure}
    \centering
    \includegraphics[width=0.5\textwidth]{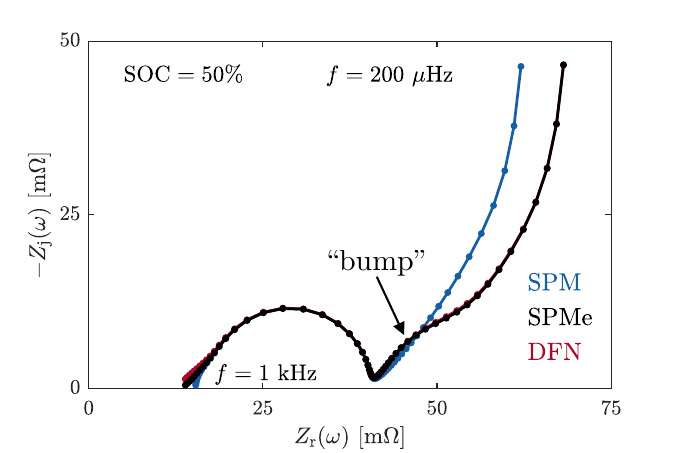}
    \caption{Nyquist plot of the EIS spectra of the SPM, SPMe, and DFN model at 50\% SOC for the LG M50 cell (\texttt{Chen2020} dataset) computed using the frequency-domain method. Frequency range: [\SI{200}{\micro Hz}, \SI{1}{kHz}]. The DFN overlaps with the SPMe, except at high frequencies. Note that we have changed the contact resistance of the SPM to match with the other models.}
    \label{Fig:Zmodels_Nyquist}
\end{figure}

Fig.~\ref{Fig:Zmodels_Nyquist} shows a Nyquist plot of the impedance for the SPM, SPMe, and DFN with double-layer capacitance at 50\%~SOC and \SI{25}{\celsius} (this is the same data as shown in Fig.~\ref{Fig:Zmethod_Bode}). For each model, we notice two main features: a semi-circle and a low-frequency diffusion tail. The semi-circle---which we also see in the measured data of Fig.~\ref{Fig:ZLGM50LTmeasured}---originates from the charge transfer kinetics interacting with the double-layer capacitance. By adding the double-layer capacitance to the model we can distinguish between the series resistance and charge transfer resistance (see Section \ref{section:ImpedanceSPMe}), which was, for instance, not the case for the SPM without double-layer capacitance \cite{bizeray2018identifiability}. An analytical form of the SPM solid-state diffusion tail can be found in Bizeray et al.\ \cite{bizeray2018identifiability}, which starts with a 45$^\circ$ slope and goes towards a capacitive behaviour (vertical line in the Nyquist plot)---we see the same here. The models where electrolyte dynamics are included (SPMe and DFN) have a diffusion tail with an additional ``bump'', linked with electrolyte diffusion. This ``bump'' is often present in measured impedance (see Fig.~\ref{Fig:ZLGM50LTmeasured}), indeed, measured data rarely shows a diffusion tail with a 45$^\circ$ slope that can be modelled with a Warburg element \cite{vandeputte2023frequency,orazem2024proper,vandeputte2024parametric}. This suggests that it is worthwhile to consider models that include electrolyte dynamics, although they are more complex. Also note that the SPMe and DFN differ only at high frequencies (which matter less for simulations of real battery behaviour), as can be observed in the Bode plot of Fig.~\ref{Fig:Zmethod_Bode}\footnote{Impedance data from the SPMe and DFN are also very similar at other SOC operating points.}.

As the SPM lacks electrolyte dynamics---which are notable in impedance measurements---and the DFN has significantly more parameters and needs longer solve times than the SPMe but shows similar impedance response, we will work with the SPMe for the remainder of this paper. 

\section{Impedance analysis of the SPMe}\label{section:ImpedanceSPMe}
We consider the SPMe with double-layer capacitance (Appendix~A), which is an extension of Marquis et al.\ \cite{marquis2019} and Brosa Planella et al.\ \cite{planella2021systematic}. This model depends on the OCPs of both electrodes and has 31 parameters\footnote{Please note that parameters with subscript $\pm$ count as two!}, listed in Table~\ref{tab:groupedParamsSPM}. Note that two of these  are physical constants (Faraday constant and ideal gas constant), and one is the measured temperature, so in total 28 unknown parameters remain. Typically, the number of necessary parameters to simulate the model can be reduced by grouping them, which is discussed next. We then study the physical processes occurring at different time-scales to show how the impedance depends on SOC and changes when varying the grouped parameters.

\subsection{Grouping model parameters}\label{section:grouping}
The model parameters cannot all be estimated independently from current and voltage data because some group together. Grouped parameters naturally arise when non-dimensionalising a model, and grouping reduces the total effective number of parameters \cite{Plett2022lumping}. This has been done for the SPM by Bizeray et al.\ \cite{bizeray2018identifiability}, for the SPM with double-layer capacitance  by Kirk et al.\ \cite{kirk2023nonlinear}, and for the SPMe without double-layer capacitance by Marquis et al.\ \cite{marquis2019}. Grouping the parameters of the DFN model has been studied in Khalik et al.\ \cite{khalik2021parameter} and in Hileman et al.\ \cite{hileman2024estimating} for lithium metal batteries. 

\begin{table}[]
    \centering
    \begin{tabular}{ll} \hline 
         & Model parameters\\ \hline
 $F$ &  Faraday constant [C/mol]\\ 
 $R_\mathrm{g}$ &Ideal gas constant [J/(mol.K)]\\
 T &Ambient temperature [K]\\
 $\alpha_\pm$&Electrode active material volume fraction \\ 
 $\varepsilon_\pm$ & Electrode porosity\\
 $\varepsilon_\mathrm{sep}$ & Separator porosity\\
 $c_{\pm,\mathrm{max}}$ & Electrode active material max.~conc. [mol/m$^3$]\\
 $L_\pm$ &Electrode thickness [m]\\
 $L$ &Total cell (electrodes \& separator) thickness [m]\\
 $A$ &Electrode area [m$^2$]\\
 $R_\pm$ &Particle radius [m] \\
 $D_{\pm}$ &Diffusivity in the particles [m$^2$/s]\\
 $D_\mathrm{e}$ &Reference electrolyte diffusivity [m$^2$/s]\\
 $C_{\mathrm{dl},\pm}$ &Electrode double-layer capacity [F/m$^2$]\\
 $m_\pm$ & Ref. exch. current dens. [(A/m$^2$)(m$^3$/mol)$^{1.5}$]\\
 $t^+$ & Cation transference number\\
 $b$ & Bruggeman coefficient\\
 $c_\pm^{*0\%}$ &Particle concentration at 0\% SOC [mol/m$^3$]\\
 $c_\pm^{*100\%}$ &Particle concentration at 100\% SOC [mol/m$^3$]\\
 $c_{\mathrm{e},0}$ &Initial electrolyte concentration [mol/m$^3$]\\
 $R_0$ & Series resistance [$\Omega$]
 \end{tabular}
 \newline
\vspace*{0.1 cm}
\newline
\renewcommand{\arraystretch}{1.7}
 \begin{tabular}{ll}
\hline
Grouped parameters $\theta$\\ 
\hline
$Q^\mathrm{th}_\pm=F\alpha_\pm c_{\pm,\mathrm{max}}L_\pm A$ &Theor. electrode capacity [As]\\ 
$\displaystyle Q_\mathrm{e}=F \varepsilon_\text{sep} c_{\mathrm{e},0} L A$ & Ref. electrolyte capacity [As] \\
$\displaystyle \tau^\mathrm{d}_\pm=\frac{R^2_\pm}{D_{\pm}}$ &Particle diff. timescale [s]\\
$\displaystyle \tau^\mathrm{e}_\pm=\frac{\varepsilon_\text{sep} L^2}{\varepsilon_\pm^{b_\pm}D_\mathrm{e}}$& Electrolyte diff. timescale [s]\\
$\displaystyle \tau^\mathrm{e}_\mathrm{sep}=\frac{L^2}{\varepsilon_\mathrm{sep}^{b_\mathrm{sep}-1}D_\mathrm{e}}$& Electrol. diff. timescale sep. [s]\\
$\displaystyle \tau^\mathrm{ct}_\pm=\frac{F R_\pm }{m_\pm \sqrt{c_{\mathrm{e},0}}}$& Charge transfer timescale [s]\\
$\displaystyle C_\pm=\frac{3\alpha_\pm C_{\mathrm{dl},\pm} L_\pm A}{R_\pm}$& Double-layer capacitance [F]\\
$\zeta_\pm = \varepsilon_\pm / \varepsilon_\mathrm{sep}$ & Relative electrode porosity \\
$\ell_\pm=L_\pm/L$ & Relative electrode thicknesses\\
$\displaystyle c_\pm^{0\%} = c_\pm^{*0\%}/c_{\pm,\mathrm{max}}$ & Stoichiometry at 0\% SOC \\
$\displaystyle c_\pm^{100\%} = c_\pm^{*100\%}/c_{\pm,\mathrm{max}}$ & Stoichiometry at 100\% SOC \\
$t^+$ & Cation transference number\\
$R_\mathrm{0}$ & Series resistance [$\Omega$]\\
\hline
\end{tabular}
\caption{Model parameters of the SPMe with double-layer capacitance detailed in Appendix~A and grouped parameters (see Appendix~C).}
\label{tab:groupedParamsSPM}
\end{table}
In Appendix~C, we group the parameters of the SPMe with double-layer capacitance, keeping only the dimensions of time, current, and voltage, and obtain 22 grouped parameters as listed in Table~\ref{tab:groupedParamsSPM}. All grouped parameters have units that depend upon seconds, Amperes, and Volts.  

The theoretical electrode capacities $Q^\mathrm{th}_\pm$ are related to the stoichiometries at 0\% and 100\% SOC through \cite{kirk2023nonlinear},
\begin{align}
        Q^\mathrm{th}_\pm=\mp \frac{Q_\mathrm{meas}}{c_\pm^{100\%}-c_\pm^{0\%}},
        \label{eq:theoreticalElectrodeCapacitySto}
\end{align}
where $Q_\mathrm{meas}$ is the measured cell capacity (e.g.\ $Q_\mathrm{meas}= \SI{5.15}{Ah} = \SI{18551}{As}$ in the \texttt{chen2020} LG~M50 parameter set). This reduces the number of grouped parameters to be estimated by two when $Q_\mathrm{meas}$ is known. 

The time-scales $\tau^\mathrm{ct}_\pm$ are used to represent the charge transfer kinetics (as these arise when non-dimensionalising the model equations). However, representing charge transfer kinetics with a resistance makes more physical sense. Typical charge transfer resistances, which have values on the order of magnitude of the diameter of the semi-circles in the impedance data, can be used instead \cite{kirk2023nonlinear}
\begin{align}
    R^\mathrm{typ}_{\mathrm{ct},\pm}&=\frac{2R_\mathrm{g} T R_\pm}{3F AL_\pm\alpha_\pm m_\pm   c_{\pm,\mathrm{max}}\sqrt{c_{\mathrm{e},0}}}=\frac{2R_\mathrm{g} T}{F}\frac{\tau^\mathrm{ct}_\pm}{3 Q^\mathrm{th}_\pm}.
    \label{eq:Rct}
\end{align}
The SPMe of Appendix~C can be simulated from the grouped parameters of Table~\ref{tab:groupedParamsSPM} and the electrode open circuit potentials (OCPs), and shows identical behaviour to the native SPMe implemented in \texttt{PyBaMM}. 

\subsection{Model time-scales}\label{section:timeScales}
\begin{figure}[htb]
    \centering
    \includegraphics[width=0.5\textwidth]{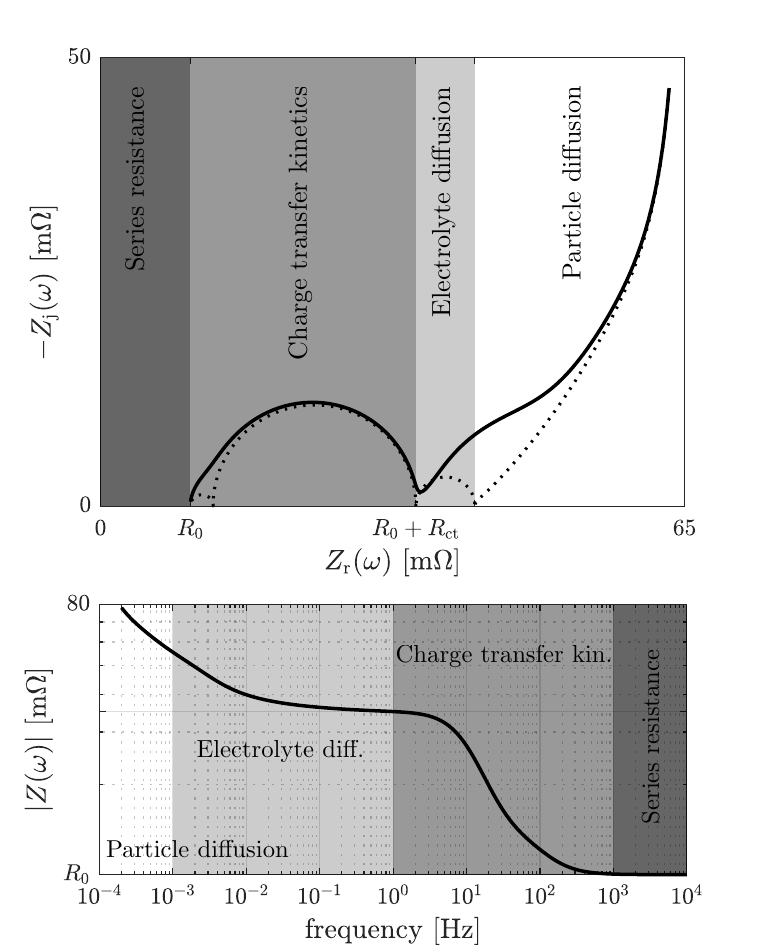}
    \caption{Impedance of the SPMe related to different physical processes at different time-scales. The dotted line in the Nyquist plot shows the impedance assuming perfect separation of time-scales whereas the black line shows the actual typical impedance.}
    \label{Fig:Zparts}
\end{figure}

Impedance data allows us to distinguish between different physical processes that occur at different time-scales \cite{lu2022timescale,ciucci2015analysis}, as illustrated for the grouped SPMe in Fig.~\ref{Fig:Zparts}. The different processes for the SPMe are the following:
\begin{itemize}
    \item $\tau<\SI{1}{ms}$ ($f>\SI{1}{kHz}$): At very short time-scales the SPMe acts as a \textbf{series resistance} $R_0$.
    \item $\SI{1}{ms} <\tau< \SI{1}{s}$ ($\SI{1}{Hz} <f< \SI{1}{kHz}$): Here, the dominant process is \textbf{charge transfer}, represented by two semi-circles in the Nyquist plot (one for each electrode). The timescale of this process is related to the charge transfer time-scales $\tau^\mathrm{ct}_\pm$. However, as discussed above, these values do not represent the actual timescale. A more representative value is $\tilde\tau^\mathrm{ct}_\pm=R^\mathrm{typ}_{\mathrm{ct},\pm}C_\pm$ (i.e., the timescale related to the corner frequency of the $RC$-pair). Numerical values for the \texttt{chen2020} parameter set are $\tilde\tau^\mathrm{ct}_+=\SI{1.5}{ms}$ and $\tilde\tau^\mathrm{ct}_-=\SI{15}{ms}$. 
    \item $\SI{1}{s} <\tau< \SI{1000}{s}$ ($\SI{1}{mHz} <f< \SI{1}{Hz}$): In this range we see \textbf{electrolyte diffusion} with time-scales $\tau^\mathrm{e}_\pm$ and $\tau^\mathrm{e}_\mathrm{sep}$. The effect on the impedance is the ``bump'' in the diffusion tail. Numerical values for the \texttt{chen2020} parameter set are $\tau^\mathrm{e}_+=\SI{409}{s}$, $\tau^\mathrm{e}_-=\SI{635}{s}$, and $\tau^\mathrm{e}_\mathrm{sep}=\SI{246}{s}$. 
    \item $\tau>\SI{1}{s}$ ($f<\SI{1}{Hz}$): The longest time-scales are related to \textbf{diffusion within the particles}, $\tau_\pm^\mathrm{d}$. The effect on the impedance is the diffusion tail. Numerical values for the \texttt{chen2020} parameter set are $\tau_+^\mathrm{d}=6812$~s and $\tau_-^\mathrm{d}=1041$~s.
\end{itemize}
The frequency ranges chosen here are not hard bounds and these values vary from battery to battery. Also note that frequency ranges of different physical processes may overlap. The electrolyte diffusion and particle diffusion time-scales overlap in this example, and so do the charge transfer time-scales in positive and negative electrodes. Due to the overlap in time-scales we get the black line in Fig.~\ref{Fig:Zparts} as the impedance, instead of the (ideal) dotted line.

\subsection{Impedance at different operating points}
\begin{figure}
    \centering
    \includegraphics[width=0.5\textwidth]{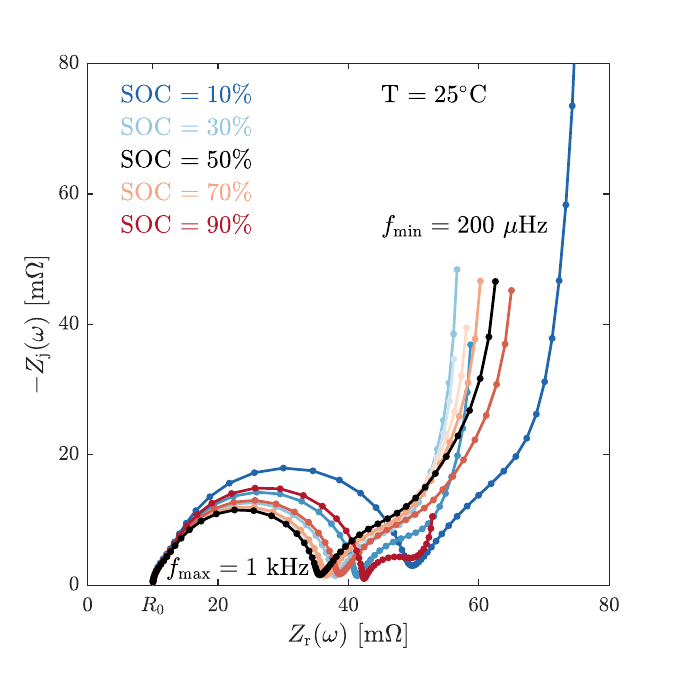}
    \caption{EIS spectra of grouped SPMe (Appendix~C) at different SOCs,    $\SI{25}{\celsius}$, with model parameters $\theta$ in Table~\ref{tab:estimationSimulation}. Frequency range: [\SI{200}{\micro Hz}, \SI{1}{kHz}]. The impedance at 10~\% SOC has been truncated.}
    \label{Fig:Zoperatingpoints}
\end{figure}

In stationary conditions, the SOC sets the particle stoichiometry $c_\pm$ (which is constant over the particle radius) by linear interpolation
\begin{align}
    c_\pm=c_\pm^{0\%}+\frac{\text{SOC}}{100}\left(c_\pm^{100\%}-c_\pm^{0\%}\right), 
    \label{eq:cpmSOC}
\end{align}
with $c_\pm^{0\%}$ and $c_\pm^{100\%}$, respectively, the stoichiometries at 0\% and 100\% SOC. Linearising the model at different SOC operating points $x_m$ (corresponding to different stoichiometries $c_{m,\pm}$) results in different impedance data, as can be seen in Fig.~\ref{Fig:Zoperatingpoints} for the grouped SPMe.

We note that the diameter of the semi-circles changes with SOC. This diameter is typically called the \textit{charge transfer resistance} and is the derivative of the voltage contribution from the spatially averaged overpotential $\overbar{\eta_\pm}$
with respect to the current, evaluated at the operating point $x_m$ (containing $i=0$). From Appendix~D,
\begin{align}
R_{\mathrm{ct},m,\pm}
&\approx \frac{R_{\mathrm{ct},\pm}^\mathrm{typ}}{2\sqrt{c_{m,\pm}(1-c_{m,\pm})}}\frac{1}{\overbar{\sqrt{c_{\mathrm{e},\pm}}}\big\rvert_{x_m}}.
\label{eq:Rctnpm}
\end{align}

High and low stoichiometries $c_{m,\pm}$, related to high and low SOC, give higher charge-transfer resistances, while the middle SOC range gives lower charge transfer resistances. This is exactly what we see from the simulated data in Fig.~\ref{Fig:Zoperatingpoints}, and can also be seen in the measured data of Fig.~\ref{Fig:ZLGM50LTmeasured}.

The diffusion tail also changes strongly with SOC, and can be explained from \eqref{eq:vpmGrouped}. The diffusion tail impedance is related to the derivative  
\begin{align}
\frac{\mathrm{d}U_\pm(c_\pm\vert_{r=1})}{\mathrm{d}i}\bigg\rvert_{x_m}=\underbrace{U_\pm'(c_{m,\pm})}_{\text{slope of OCP}}\frac{\mathrm{d}c_\pm\vert_{r=1}}{\mathrm{d}i}\bigg\rvert_{x_m},
\label{eq:slopeDiffusionTail}
\end{align}
and is proportional to the slope of the OCP \cite{kirk2023nonlinear,bizeray2018identifiability}. This strongly depends on the stoichiometries $c_{m,\pm}$, and, hence, on the SOC (see Fig.~\ref{Fig:OCPLGM50LT}). The OCP slopes here are largest at 10\%~SOC, giving the largest diffusion tail. At an SOC where an OCP has zero slope there will be no contribution from this electrode in the diffusion tail.

The series resistance $R_0$ in this model does not depend on SOC, and hence, the leftmost intercept with the real axis stays the same for all SOC.

\subsection{Impedance sensitivity to grouped parameters}
\begin{figure*}
    \centering
    \includegraphics[width=1\textwidth]{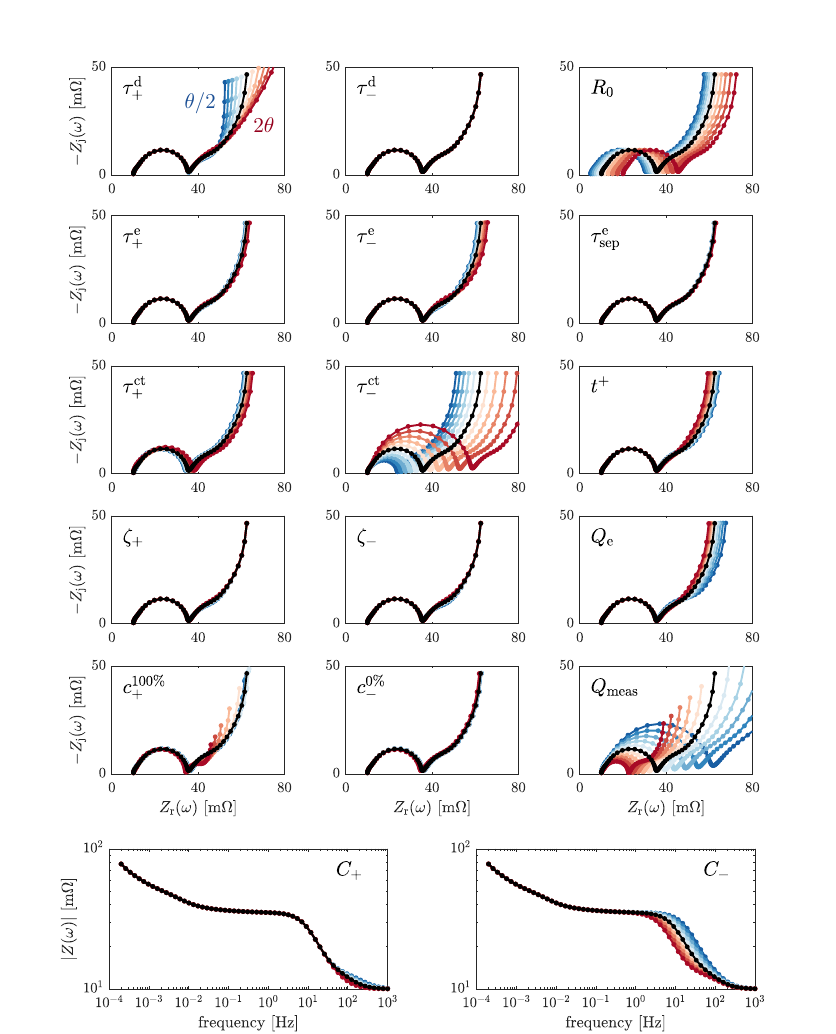}
    \caption{Impedance of SPMe at 50\% SOC, $\SI{25}{\celsius}$, varying  grouped parameters one-by-one. Parameters are perturbed with logarithmic spacing in the range $[0.5\theta,2\theta]$ (blue to red) with $\theta$ listed in Table~\ref{tab:estimationSimulation} (black). Frequency range: [\SI{200}{\micro Hz} , \SI{1}{kHz}]. Note that most impedance are shown on Nyquist plots, while the bottom two figures are Bode magnitude plots.}
    \label{Fig:Zsensitivity}
\end{figure*}

Fig.~\ref{Fig:Zsensitivity} shows the impedance of the SPMe for variations around the grouped parameters $\theta$ of Table~\ref{tab:estimationSimulation}:
\begin{itemize}
\item [$\tau^\mathrm{d}_\pm$] Particle diffusion time-scales have an effect at low frequencies in the diffusion tail, associated with the frequency where the diffusion tail changes from the approx.\ 45 degree slope to capacitative behaviour. Long experiments are needed to estimate these parameters. At 50\% SOC $\tau^\mathrm{d}_-$ seems not to affect the impedance, because the slope of the negative OCP is nearly flat at that SOC \cite{bizeray2018identifiability}. At other SOCs the diffusion timescale of different electrodes may be more identifiable, as can be seen in Fig.~\ref{Fig:Ztausensitivity}. This motivates estimating model parameters from impedance at several SOCs, as in \eqref{eq:costfunctionImpedance}. 

\item [$R_0$] Series resistance simply shifts the impedance horizontally.

\item [$\tau^\mathrm{e}$] Electrolyte diffusion time-scales impact the timescale of the ``bump'', and hence, at which frequencies it interacts with the particle diffusion. The electrolyte diffusion timescale in the separator $\tau^\mathrm{e}_\mathrm{sep}$ has almost no effect on the impedance.

\item [$\tau^\mathrm{ct}_\pm$] Charge transfer time-scales change the diameters of the semi-circles. The change for the negative electrode is observed to be significantly larger than for the positive one. The diameter of the semi-circles is larger around 0 and 100\% SOC, as can be seen in Fig.~\ref{Fig:Zoperatingpoints}, making it easier to estimate these parameters. This again supports estimating model parameters from impedance at different SOC levels. 

\item [$t^+$] Cation transference number slightly changes the initial slope and resistance of the diffusion tail. 

\item [$\zeta_\pm$] Relative electrode porosities have a small effect in the middle part of the diffusion tail.

\item [$Q_\mathrm{e}$] Reference electrolyte capacity changes the initial slope and resistance of the diffusion tail.

\item [$c_\pm^{0\%}$] Stoichiometries at 0 and 100\%~SOC change the stoichiometry at the particular SOC \eqref{eq:cpmSOC}, and, hence, the width of the semi-circle \eqref{eq:Rctnpm} and the diffusion tail \eqref{eq:slopeDiffusionTail}.

\item[$Q_\mathrm{meas}$] Measured capacity changes the total size of the impedance; large capacity implies small impedance and vice versa. The theoretical electrode capacities $Q_\pm^\mathrm{th}$ can be calculated from $Q_\mathrm{meas}$, $c_\pm^{0\%}$, and $c_\pm^{100\%}$ \eqref{eq:theoreticalElectrodeCapacitySto}.

\item [$C_\pm$] The impact of double-layer capacitance cannot easily be seen on Nyquist charts, because it only changes the frequency dependence of the points on the semi-circle, but it can be seen on a magnitude Bode plot. Here we note that the charge-transfer kinetics at the surface of the positive particles are faster than the negative ones. 
\end{itemize}

As all grouped parameters have a unique effect on the impedance, they can possibly be estimated from impedance data (depending on the frequency range and how strongly the processes are present in the impedance).

\section{Parametrisation of the SPMe from simulations}
We now study estimation of grouped SPMe parameters from impedance data by fitting the model to simulated data where true parameters are known.

The grouped parameters $c_\pm^{0\%}$ and $c_\pm^{100\%}$ are typically estimated from OCP and OCV data \cite{kirk2023nonlinear} prior to impedance fitting. However, instead of fixing these from OCP and OCV data, we will estimate them from impedance data at different SOCs, and then set $Q^\mathrm{th}_\pm$ from \eqref{eq:theoreticalElectrodeCapacitySto}. 

Assuming that the relative thicknesses $\ell_\pm$ are known, the remaining grouped model parameters to estimate from impedance data are
\begin{align}
    \theta = [&\tau^\mathrm{d}_+,\tau^\mathrm{e}_+,\tau^\mathrm{ct}_+,C_+,\zeta_+,c_+^{0\%},c_+^{100\%},\nonumber\\&\tau^\mathrm{d}_-,\tau^\mathrm{e}_-,\tau^\mathrm{ct}_-,C_-,\zeta_-,c_-^{0\%},c_-^{100\%},\nonumber\\&\tau^\mathrm{e}_\mathrm{sep},Q_\mathrm{e},t^+,R_0]\qquad\in\mathbb{R}_+^{18\times 1}.
    \label{eq:thetaSPMe}
\end{align}
The values of the true grouped parameters $\theta$ for the simulations are calculated from the \texttt{Chen2020} parameter set and listed in Table~\ref{tab:estimationSimulation}.

\begin{table*}[htb]
\centering
\begin{tabular}{lcl|ll|ll} 
 & & & Impedance data& & Voltage data& \\
\hline
 Parameter &  Optimisation bounds & $\theta$ & $\hat \theta$ & $\sigma_{\hat \theta}/\hat \theta$ [\%]& $\hat \theta$ & $\sigma_{\hat \theta}/\hat \theta$ [\%]\\
\hline
$Q_\mathrm{meas}$ [As] & & 18551& & & &\\
$l_+$ & &0.4375& & & &\\
$l_-$ & &0.4930& & & &\\ 
\hline
$\tau^\mathrm{d}_+$ [s]&[5e2,1e4]&6812& 6682& 0.38& 6831& 0.53\\
$\tau^\mathrm{d}_-$ [s]&[5e2,1e4]&1041& 1020& 0.93& 1065& 1.48\\
\hline
$\tau^\mathrm{e}_+$ [s]&[2e2,1e3]& 409.2& 200.0&  0.011& 948.0& 1.24\\
$\tau^\mathrm{e}_-$ [s]&[2e2,1e3]& 634.7& 403.5& 20.9& 209.9& 53.7\\
$\tau^\mathrm{e}_\mathrm{sep}$ [s]&[2e2,1e3]&246.2& 200.0& 0.61& 200.7& 0.59\\
\hline
$\zeta_+$ & [0.5,1.5]&  0.7128& 0.500& 0.0019& 0.7573& 3.13\\
$\zeta_-$ & [0.5,1.5]& 0.5319& 1.500& 41.9& 0.5420& 3.09\\
$Q_\mathrm{e}$ [As] & [5e2,1e3]& 804.8& 500.0& 8.20& 797.9& 2.12\\
\hline
$\tau^\mathrm{ct}_+$ [s] &[1e3,5e4]&4657& 5158& 26.5& 6244& 6.92\\
$\tau^\mathrm{ct}_-$ [s] &[1e3,5e4]&27592& 27746& 5.07& 13255& 8.38\\
\hline
$C_+$ [F]&[0,1]& 0.5935& 0.6178& 13.9& 1.2e-4& 167\\ 
$C_-$ [F]&[0,1]& 0.6719& 0.6953& 8.68& 0.9966& 4.04\\ 
\hline
$c^{0\%}_+$ & [0.8,0.9]& 0.8540& 0.8548& 0.10& --& --\\
$c^{0\%}_-$ & [0,0.1]& 0.02635& 0.02634& 11.8& --& --\\
$c^{100\%}_+$ & [0.2,0.3]& 0.2638& 0.2642& 0.22& --& --\\
$c^{100\%}_-$ & [0.85,0.95]& 0.9106& 0.9092& 0.31& --& --\\
\hline
$t^+$ &[0.2,0.5]&0.2594 &  0.2647& 9.16& 0.2025& 28.7\\
$R_0$ [$\Omega$]&[0,0.05]&0.01& 0.0101& 1.22& 0.0148& 8.56\\
\hline
 \end{tabular}
 \newline
\vspace*{0.2 cm}
\newline
\begin{tabular}{llllllllll} \hline 
$m$ &  1& 2&  3& 4& 5 & 6& 7& 8& 9 \\  
SOC$_m$ [\%] &  10& 20&  30& 40& 50& 60& 70&  80& 90\\ 
FE$_m$ [\%] & 0.87&  0.97& 0.96& 0.97& 0.94& 0.94& 0.90& 0.86& 0.84\\ \hline
\end{tabular}
\caption{Parametrisation of the grouped SPMe from simulated impedance and time-domain voltage data with particle swarm optimisation in \texttt{PyBOP}. We list the optimisation bounds, true parameters $\theta$, estimated parameters $\hat \theta$ (lowest cost of ten runs), and relative standard deviation $\sigma_{\hat \theta}/\hat \theta$ (over ten runs). The impedance data is shown in Fig.~\ref{Fig:Zoperatingpoints} and the time-domain voltage data in Fig.~\ref{Fig:timeDomainSimulation}.
The average computation times (run parallel on 16 AMD 7950X cores) are \SI{441}{s} (impedance) and \SI{613}{s} (time-domain voltage). Better estimates for the electrolyte diffusion parameters can be obtained using other optimisers at the cost of longer computation times. The fitting errors \eqref{eq:fittingError} of the impedance at different SOC are also listed.} 
\label{tab:estimationSimulation}
\end{table*}

\subsection{Estimation from impedance data}
\begin{figure}[htb]
    \centering
    \includegraphics[width=0.5\textwidth]{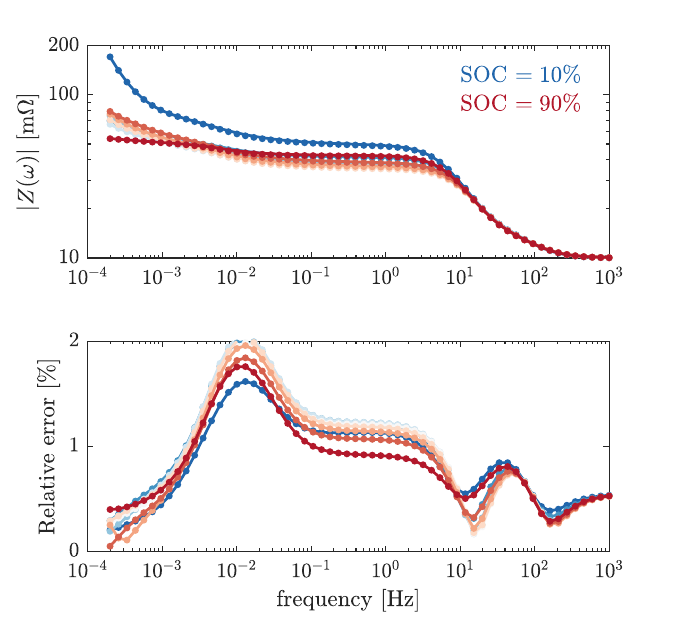}
    \caption{Fits of simulated SPMe impedance data and relative error. The estimated parameters are listed in Table~\ref{tab:estimationSimulation}.}
    \label{Fig:errorSimulation}
\end{figure}

We first study parameter estimation from simulated SPMe impedance data (shown in Fig.~\ref{Fig:Zoperatingpoints}). The impedance dataset \eqref{eq:impedanceDataset} has $K=60$ frequencies, logarithmically spaced between \SI{200}{\micro Hz} and \SI{1}{kHz}, and $M=9$ SOC levels (10\%, 20\%,..., 90\%).  We estimate the grouped parameters $\theta$ \eqref{eq:thetaSPMe} by minimising the least-squares cost function \eqref{eq:costfunctionImpedance}. The optimisation is performed with the multi-fitting problem class in \texttt{PyBOP}, using particle swarm optimisation, with the boundaries for the 18-dimensional search space given in Table~\ref{tab:estimationSimulation}. We allowed for a maximum of 1000 iterations and ran the optimisation ten times. Although the fitting is done over all SOCs, it is instructive to examine the mean relative fitting errors (FE) at the different SOC levels
\begin{align}
    \mathrm{FE}_m 
\ [\%]=\frac{100}{K}\sum_{k=1}^{K}\frac{\vert Z_m(\omega_k)-Z_m(\omega_k,\hat \theta)\vert}{\vert Z_m(\omega_k)\vert}.
\label{eq:fittingError}
\end{align}
These are listed in Table~\ref{tab:estimationSimulation} for the best fit of the ten runs and the data over frequency is shown in Fig.~\ref{Fig:errorSimulation}. The fitting errors are under 1\% for all SOC levels. 

The estimated parameters $\hat \theta$ of the best fit of the ten runs are listed in Table~\ref{tab:estimationSimulation}. As an indication of the identifiability of the grouped parameters, we also list the relative standard deviation $\sigma_{\hat \theta}/\hat \theta$ over the ten runs; we would expect accurately identifiable parameters to have a small standard deviation. However, the cost function is highly non-convex, and, hence, this metric for the identifiability may depend on the optimiser and initial point. (A better way would be to consider the Fisher information matrix \cite{rissanen1996fisher}.) The parameters related to large features can be identified accurately from impedance data (e.g.\ the stoichiometry bounds, series resistance, particle diffusion time-scales, and charge transfer timescale). The estimated parameters related to the electrolyte diffusion and separator are less accurate, but these parameters related to small features in the impedance have little effect on the behaviour of the battery. The grouped model parameters are also all attributed to the correct electrode. 

\subsection{Comparison with time-domain data}
As a comparison, parameters were also estimated from the simulated SPMe time-domain voltage data of Fig.~\ref{Fig:timeDomainSimulation}. The dataset consists of 7 minutes resting at 90\% SOC, starting from steady-state, then a \SI{-5}{A} discharge for 53 minutes, 20 minutes rest, \SI{5}{A} charge for 20 minutes, and 20 minutes rest again, resulting in a total experiment time of 2~h. The data were simulated with a time step of \SI{10}{s}, resulting in $N=720$ points. Note that measuring this dataset would take less time than measuring the impedance dataset from Fig.~\ref{Fig:Zoperatingpoints}. We estimate the grouped parameters $\theta$ \eqref{eq:thetaSPMe} by minimising the least-squares cost function \eqref{eq:timeDomainOptimisation}. The fitting was performed in \texttt{PyBOP}, using particle swarm optimisation, with the same boundaries for the 18-dimensional search space given in Table~\ref{tab:estimationSimulation}. We allow for a maximum of 1000 iterations and run the optimisation ten times. The stoichiometries $c_\pm^{0\%}$ and $c_\pm^{100\%}$ were not estimated as this led to convergence issues; they were assumed known.

Accurate fits of the time-domain voltage data were obtained, with the estimated parameters of the best fit and the relative standard deviations over the ten runs listed in Table~\ref{tab:estimationSimulation}. For this specific simulation, we conclude that the long timescale parameters are identified with similar accuracy from the time-domain data and impedance data, while the short timescale parameters are better identified from impedance data because the sampling period of \SI{10}{s} in the voltage time-domain data does not give visibility of the fast processes occurring in the battery. 

The comparison of the different sources of data discussed above suggests an extension of the work presented here where one could use both time-domain voltage and EIS data to estimate model parameters.

\section{Parametrisation of LG M50LT batteries from measured data}
\begin{figure}
    \centering
    \includegraphics[width=0.5\textwidth]{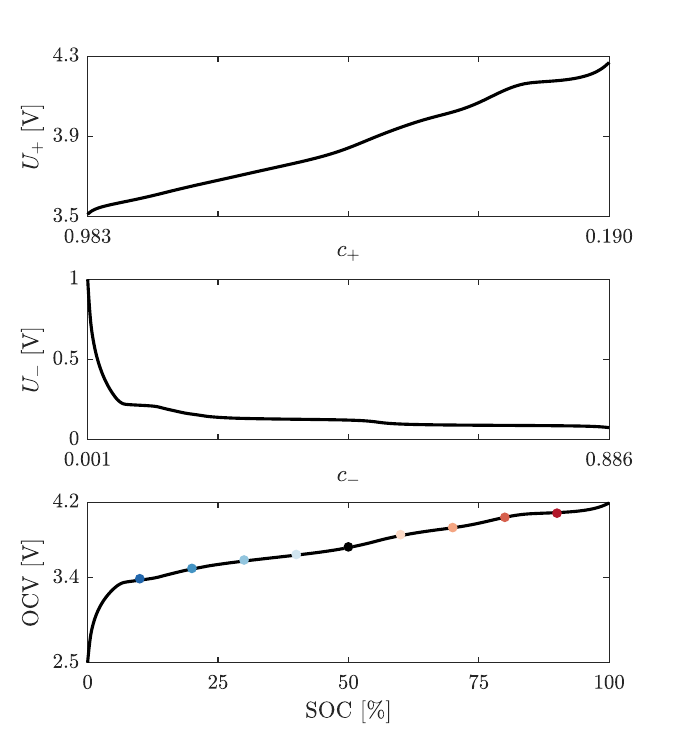}
    \caption{Measured OCV and OCP data for the LG M50LT. Dots indicate points where the impedance spectra of Fig.~\ref{Fig:ZLGM50LTmeasured} were measured.}
    \label{Fig:OCPLGM50LT}
\end{figure}

\begin{figure}[htb]
    \centering
    \includegraphics[width=0.5\textwidth]{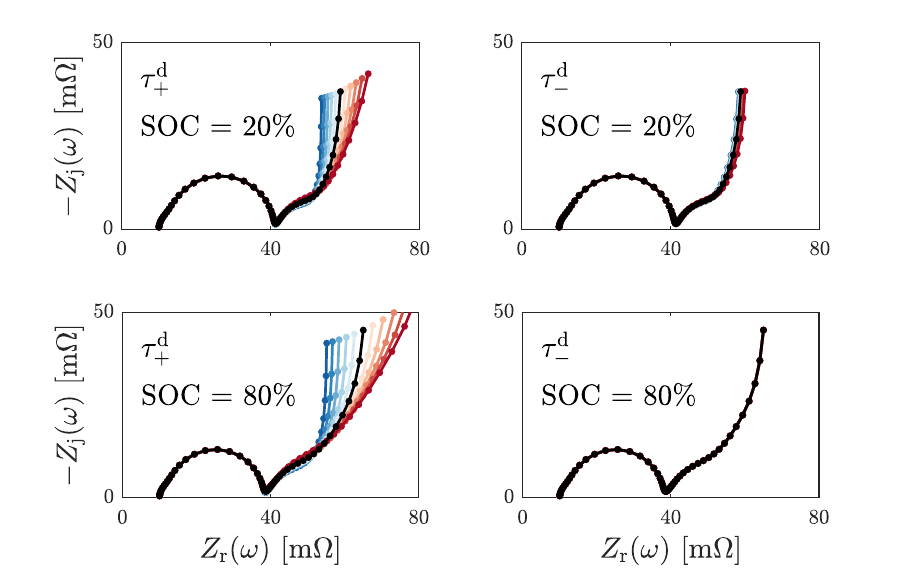}
    \caption{Sensitivity of the SPMe impedance to particle diffusion time-scales $\tau_\pm^\mathrm{d}$. The impedance is insensitive to $\tau_\pm^\mathrm{d}$ at SOC where the OCP $U_\pm$ is flat. The OCP of graphite has several plateaus (see Fig.~\ref{Fig:OCPLGM50LT}), which explains why $\tau_-^\mathrm{d}$ is not identifiable at certain SOC. Parameters are perturbed in the range $[0.5\theta,2\theta]$ (blue to red) with $\theta$ the Chen2020 parameters with contact resistance of 10~m$\Omega$ (black). Frequency range: [200 $\mu$Hz, 1 kHz].}
    \label{Fig:Ztausensitivity}
\end{figure}

We now parametrise the grouped SPMe from measured LG M50LT impedance data. This Li-ion cell has a graphite anode and NMC cathode, and operates between \SI{2.5}{V} and \SI{4.2}{V}. The measured cell capacity was $Q_\mathrm{meas}=\SI{4.89}{Ah}$. Open circuit potentials were obtained by constructing half cells; this data was provided by About:Energy (personal communication) and is shown in Fig.~\ref{Fig:OCPLGM50LT} for the given stoichiometric bounds. 

\begin{table*}[htb]
\centering
\begin{tabular}{lccccccccc} \hline 
$m$ &  1& 2&  3& 4& 5& 6& 7& 8& 9\\ 
SOC$_m$ [\%] &  10& 20&  30& 40& 50& 60& 70& 80& 90\\ 
OCV$_m$ [V] & 3.391  &  3.500  &  3.589 & 3.647 & 3.728 & 3.858 & 3.933 & 4.041 & 4.085\\
SNLDR$_m$& 246& 1061& 702& 558& 303& 82 & 118& 267& 579\\
$\mathrm{FE}_m$ [\%] & -- & 2.57& 1.66& 3.25& 3.58& 2.44& 3.56& 3.09& -- \\
\hline
\end{tabular}
\caption{The operating points for the LG~M50LT cell that were considered here with the signal-to-nonlinear distortion ratio (SNLDR) \eqref{eq:SNLDR} at \SI{400}{\micro Hz}. Ambient temperature throughout was \SI{25}{\celsius}.}
\label{Table:operatingPointsMeasurementsLGM50LT}
\end{table*}
\begin{figure}
    \centering
    \includegraphics[width=0.5\textwidth]{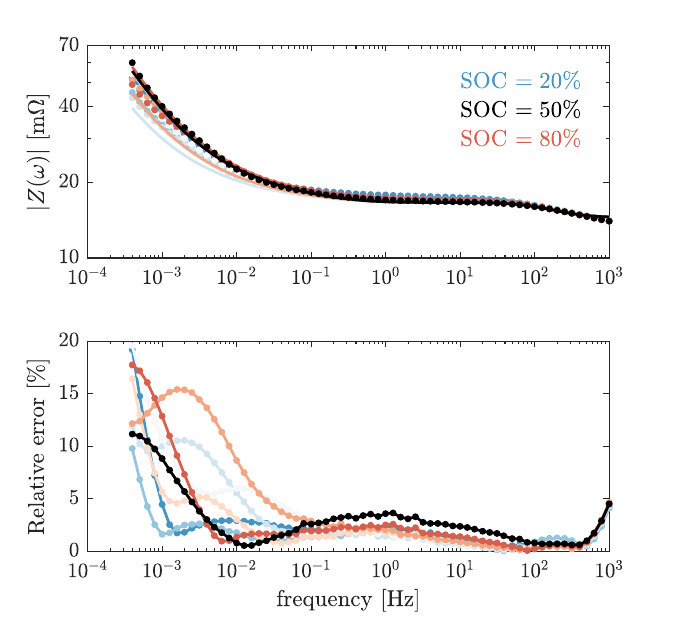}
    \caption{SPMe fits to impedance measurements LG~M50LT battery.}
    \label{Fig:MeasurementFits}
\end{figure}
\begin{table}
    \centering
    \begin{tabular}{lll}
    \hline
    $\tau^\mathrm{d}_+=8953$~s& $\tau^\mathrm{d}_-=7974$~s & $R_0=14.5$~m$\Omega$\\
    $\tau^\mathrm{e}_+=100^*$~s& $\tau^\mathrm{e}_-=100^*$~s & $\tau^\mathrm{e}_\mathrm{sep}=100^*$~s\\
    $\zeta_+=1.70^*$ & $\zeta_-=1.70^*$ &$Q_\mathrm{e}=2000^*$~As\\
    $\tau^\mathrm{ct}_+=1087$~s & $\tau^\mathrm{ct}_-=1515$~s  & $t^+=0.70^*$ \\
    $C_+=2^*$~F& $C_-=0.532$~F& $l_+=0.370$\\ 
    $c^{0\%}_+=0.983$ & $c^{0\%}_-=0.0015$& $l_-=0.551$\\
    $c^{100\%}_+=0.190$ & $c^{100\%}_-=0.886$& $Q_\mathrm{meas}=17586$~As\\
    \hline
    \end{tabular}
    \caption{Grouped parameters for the LG~M50LT cell. Parameters indicated with an asterisk have clipped at their optimisation bounds.}
    \label{tab:paramsLGM50LT}
\end{table}
\subsection{Impedance measurements}
We performed EIS measurements at different operating points, listed in Table~\ref{Table:operatingPointsMeasurementsLGM50LT} and shown as dots in Fig.~\ref{Fig:OCPLGM50LT}, with a Gamry Interface 5000P potentiostat. The battery was first charged to \SI{4.2}{V} and held there until the current decreased below \SI{50}{mA}. It was then discharged at C/10 in steps of 10\% SOC (with $Q_\mathrm{meas}/10=\SI{0.489}{A}$ for periods of 1 hour), followed by 4 hours of open-circuit relaxation to reach steady-state for the EIS measurement. Impedance was measured in the frequency range [\SI{400}{\micro Hz}, \SI{1}{kHz}] with ten logarithmically distributed frequencies per decade (65 frequencies in total). Hybrid EIS was used (a current is applied, but a voltage perturbation is chosen by the user) with a DC current of \SI{0}{A} and a sinusoidal voltage perturbation of \SI{3}{mV\ rms} ($\approx$\SI{4.2}{mV} amplitude). A Nyquist plot of the measured impedance data is shown in Fig.~\ref{Fig:ZLGM50LTmeasured}. 

The conditions of linearity and stationarity were checked from the raw current and voltage signals provided by the potentiostat by plotting their Lissajous curves and the discrete Fourier transform. The second harmonic was dominant, while higher harmonics were mostly at noise level. We define the signal-to-nonlinear distortion ratio (SNLDR) as 
\begin{align}
    \text{SNLDR}=\vert V(\omega)/V(2\omega)\vert \qquad \text{for }\omega = 2\pi f_\text{min},
    \label{eq:SNLDR}
\end{align}
which we evaluate for the different SOC levels at the lowest frequency (\SI{400}{\micro Hz}), as the measurements are most prone to be nonlinear at low frequencies due to the nonlinear OCV. The SNLDR is high at most SOC levels (see Table~\ref{Table:operatingPointsMeasurementsLGM50LT}), with lowest value of 82 at 60\% SOC (around which the OCV is quite nonlinear, see Fig.~\ref{Fig:OCPLGM50LT}). We conclude that the SNLDR over all SOC is large enough for our application and the EIS data can be considered as linear and stationary.

\subsection{Parameter estimation}
The grouped SPMe parameters $\theta$ \eqref{eq:thetaSPMe} were estimated from impedance data between 20\% and 80\% SOC (the 10\% and 90\% SOC data were hard to fit with the SPMe).  The cost function \eqref{eq:costfunctionImpedance}, with $M=7$ and $K=65$, was minimised with 1000 iterations of PSO. The estimated grouped parameters are listed in Table~\ref{tab:paramsLGM50LT}, with fits shown in Fig.~\ref{Fig:MeasurementFits}, and average fitting errors listed in Table~\ref{Table:operatingPointsMeasurementsLGM50LT}. Fits at (very) low frequencies are relatively poor, but for higher frequencies we obtained better results. At low frequencies the impedance is strongly dependent on the slope of the OCPs and due to hysteresis the OCPs may be dependent on the direction from which the operating point is approached (charging or discharging),  making it hard to obtain good fits. Also, the SPMe may not be general enough to accurately fit measured LG~M50LT data. Note that some estimated parameters in Table~\ref{tab:paramsLGM50LT} have clipped at their optimisation bounds (indicated with an asterisk). These parameters are related to small features in the impedance and are difficult to estimate, especially as the model has not enough freedom to fit the low-frequency data well. Changing the optimisation bounds did not solve this issue.

We have not explored fitting the measured data with other models but there are many possible extensions (which can be implemented in \texttt{PyBaMM}) that might improve the fit. These include considering the diffusivities in the particles to change with SOC (for example based on OCP gradients \cite{mendoza2016, horner2021}), or considering a distribution of particle sizes using multi-particle models \cite{kirk2022modeling}, as well as the many other models available in \texttt{PyBaMM}. 

\subsection{Model validation on drive cycle}
\begin{figure}
    \centering
    \includegraphics[width=0.5\textwidth]{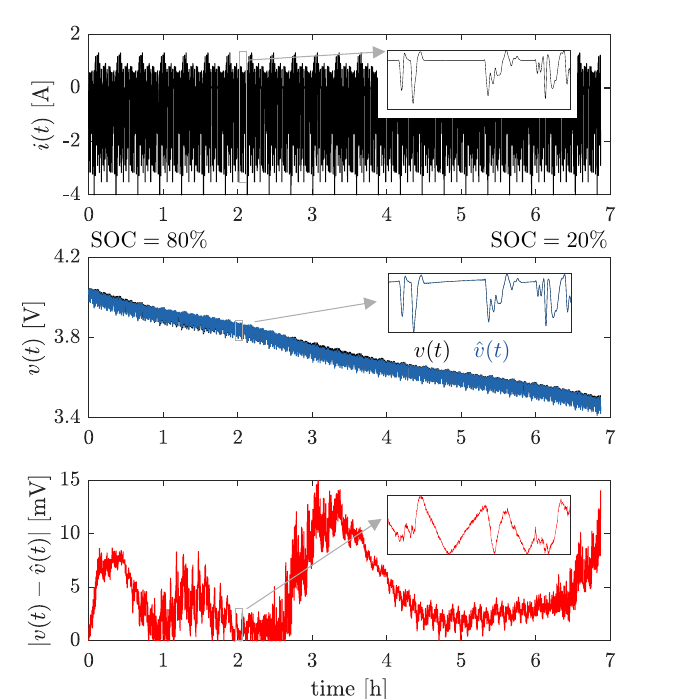}
    \caption{Validation of the parametrised SPMe using a drive cycle between 80 and 20\% SOC. We show the measured voltage $v(t)$ (black---hidden beneath blue), simulated voltage $\hat v(t)$ of the SPMe of Appendix~C with model parameters listed in Table~\ref{tab:paramsLGM50LT} (blue), and the absolute difference between the two. The boxes show zoomed plots starting at $t=2$~h.}
    \label{Fig:ValidationSPMeLGM50LT}
\end{figure}
The parametrised SPMe from LG~M50LT impedance data is now validated on a drive cycle to show its performance predicting voltage  (see~Fig.~\ref{Fig:ValidationSPMeLGM50LT}). The same drive cycle as in Kirk et al.\ \cite{kirk2023nonlinear} (with largest discharge current \SI{4}{A} and largest charge current \SI{1}{A}) is consecutively repeated 23.41 times to cover the range from 80 to 20\% SOC, resulting in a measurement time of \SI{6}{h} \SI{52}{min}. The drive cycle is sampled with a period of \SI{0.15}{s}. The voltage response of the LG~M50LT $v(t)$ is measured and the voltage response of the grouped parameter SPMe (Appendix~C and parameters listed in Table~\ref{tab:paramsLGM50LT}) $\hat v(t)$ is simulated. The absolute difference between the two is shown in red, with a maximal error of \SI{15}{mV} and root-mean-square error of \SI{5.3}{mV}. Despite the low frequency impedance fits being relatively poor (see Fig.\ \ref{Fig:MeasurementFits}), these voltage errors are reasonable for several applications of the model.

\section{Conclusions} 
We have proposed a numerical approach to compute the impedance of physics-based battery models, implemented under \texttt{PyBaMM-EIS}. The method consists of discretising the model in space, locally linearising the resulting nonlinear system of DAEs using automatic differentiation, transforming the linearised equations into the frequency-domain, and solving for the impedance. This tool allows one to evaluate the impedance of \textit{any} \texttt{PyBaMM} model at \textit{any} operating point, and is significantly faster than doing ``brute-force'' simulations while giving comparable results. 

We have analysed the impedance of different physics-based models (SPM, SPMe, and DFN), and shown that it is essential to consider electrolyte dynamics, which add a ``bump'' in the diffusion tail that is also present in measured data. The DFN has similar impedance to the SPMe, but is more expensive to simulate and has more parameters. Therefore, we conclude that most of the behaviour of a battery can be explained via the impedance of the SPMe.

We have provided the model equations for the SPMe including double-layer capacitance, and grouped the parameters of this model to determine a minimal set of parameters required for simulations. The grouped-parameter SPMe gives identical results to \texttt{PyBaMM}'s SPMe. We also showed how the impedance depends on SOC and, via a sensitivity analysis, how the grouped parameters affect the impedance in different ways.

We estimated 18 SPMe grouped parameters using simulated impedance data and voltage data. Crucially, we fitted impedance data across different SOCs simultaneously in order to increase the identifiability of the parameters. Simulated EIS data could be fitted accurately, with most grouped parameters well estimated. The exceptions were parameters related to small features in the impedance, which have little effect on the overall behaviour, for example the parameters of the separator. Hence, EIS can be used as an effective tool to parametrise the SPMe. Comparing the informativity of simulated impedance vs.\ time-domain voltage data, we found that the long timescale parameters (e.g.\ particle diffusion time-scales) can be estimated accurately from both, but the short timescale parameters are better estimated from impedance data. We therefore suggest exploring both sources of data for parameter estimation.

To demonstrate how the methodology can be used in practice, we performed parametrisation of the SPMe for a commercial LG~M50LT battery from measured impedance and OCP data. The fit was good in the higher frequency region, but less good at lower frequencies. This suggests that the SPMe needs extensions to accurately describe battery behaviour (e.g. particle size distributions and functional parameters). Still, the parametrised SPMe performed well on a drive cycle validation.

The methodology presented in this paper gives a practical method for fitting a large class of battery models to EIS data. The use of automatic differentiation combined with working in the frequency-domain dramatically reduces the computation time, and fitting the parameters across many SOC levels gives accurate estimations for most parameters. 

\section*{Appendix A: Dimensional SPMe}
\setcounter{equation}{0}
\renewcommand{\theequation}{A.\arabic{equation}}

Here, we detail the dimensional SPMe model based on \cite[Chapter 3]{marquisthesis} with improved averaging from \cite{planella2021systematic} and double-layer capacitance from \cite{kirk2023nonlinear}. The variables in this model are listed in Table~\ref{tab:SPMeVariables} and the parameters in Table~\ref{tab:groupedParamsSPM}. In this model we choose a charging current to be positive (such that the impedance ends up in the right quadrant of the complex plane). Note that this is different to the implementations in \texttt{PyBaMM} and \texttt{PyBOP}, where charging currents are negative (and a corresponding minus sign is included in calculations of the impedance).

\begin{table}[htb]
\centering
\begin{tabular}{cl} \hline 
     & Model variables \\ \hline
 $i(t)$ & Applied current [A]\\
 $v(t)$ & Terminal voltage [V]\\
 $\overbar{v_\pm}(t)$ & Electrode-aver. particle surface volt. [V]\\
 $U_\pm(c_\pm)$ & Open circuit voltage [V]\\
 $c_\pm(r,t)$ & Particle lithium concentration [mol/m$^3$]\\
 $j_\pm(x,t)$ & Molar flux [mol/(m$^2$s)]\\
 $\eta_\pm(x,t)$ & Overpotentials [V]\\
 $c_\mathrm{e}(x,t)$ & Electrolyte lithium concentr. [mol/m$^3$]\\
 $N_\mathrm{e}(x,t)$ & Electrolyte flux [mol/(m$^2$s)]\\
 $i_{0,\pm}(x,t)$ & Ref. exchange current density [A/m$^2$]\\
 $r$ & Particle radius [m]\\
 $x$ & Electrode thickness [m]\\
 $t$ & Time [s]\\
 \hline
 \end{tabular}
 \caption{Variables of the SPMe with their units.}
 \label{tab:SPMeVariables}
 \end{table}
\paragraph{Electrode average operator}
For any variable $h$, we define an ``electrode-average'' operator for each electrode, denoted by an overbar and subscript ($+/-$), as
\begin{align}
&\overbar{h_\pm}(t)=\frac{1}{L_{\pm}}\int h(x,t) \;\mathrm{d}x_\pm, 
\end{align}
with the integration domains  $\mathrm{d}x_-\in[0,L_-]$
 and $\mathrm{d}x_+\in[L-L_+, L]$.
 
\paragraph{Diffusion in the spherical solid particles in each of the electrodes ($\pm$)}
\begin{align}
\frac{\partial c_{\pm}}{\partial t} = \frac{1}{r^2}\frac{\partial}{\partial r}\left(r^2 D_{\pm}(c_{\pm})\frac{\partial c_{\pm}}{\partial r} \right) \qquad 0\leq r\leq R_\pm
\end{align}
with boundary conditions
\begin{align}
\left. D_{\pm}(c_{\pm})\frac{\partial c_{\pm}}{\partial r}\right\vert_{r=0}= 0, \qquad \left. -D_{\pm}(c_{\pm})\frac{\partial c_{\pm}}{\partial r}\right\vert_{r=R_\pm}= \overbar{j_{\pm}},
\end{align}
and initial conditions 
\begin{align}
    c_{\pm}(r,0)=c_\pm^{0\%}+\frac{\text{SOC}_0}{100}\left(c_\pm^{100\%}-c_\pm^{0\%}\right).
\end{align}

\paragraph{Intercalation reaction at the particle surface}
\begin{align}
     j_{\pm}(x,t)=
\frac{2 i_{0,\pm}}{F}\sinh\left(\frac{F\eta_\pm}{2R_\mathrm{g}T}\right)
\label{eq:overpotential}
\end{align}
with the reference exchange current density
\begin{align}
    &i_{0,\pm}(x.t)=m_{\pm}\sqrt{c_\pm |_{r=R_{\pm}}} \sqrt{c_{\mathrm{e},\pm}(c_{\pm,\mathrm{max}}-c_{\pm} |_{r=R_\pm})},
\end{align}
the overpotentials
\begin{align}
    \eta_\pm(x,t) &= v_{\pm} -U_{\pm}(c_{\pm}|_{r=R_{\pm}}),
\end{align}
and the particle surface voltage
\begin{align}
    v_\pm(x,t) = \overbar{v_{\pm}} +\frac{2R_\mathrm{g}T}{F}(1 - t^+)\left(\overbar{\log(\frac{c_{\mathrm{e},\pm}}{c_{\mathrm{e},0}})} - \log(\frac{c_{\mathrm{e},\pm}}{c_{\mathrm{e},0}})\right).
\end{align}

\paragraph{Transfer process at the particle surface with double-layer}  
\begin{align}
C_{\mathrm{dl},\pm}\frac{\mathrm{d} \overbar{v_{\pm}}}{\mathrm{d}t} = \pm\frac{R_{\pm}}{3 \alpha_\pm L_{\pm}A}i(t) - F\overbar{j_{\pm}}
\end{align}
with initial condition $\overbar{v_{\pm}}(0)=U_{\pm}(c_{\pm, 0})$.

\paragraph{Diffusion in the electrolyte} 
\begin{align}
\varepsilon(x) \frac{\partial c_\mathrm{e}}{\partial t} = &-\frac{\partial N_\mathrm{e}}{\partial x} \nonumber\\
&+
\left\{
\begin{tabular}{ll}
$\displaystyle{\frac{3\alpha_-}{R_-}j_-}$ & for  $0<x<L_-$ \\ 
$0$ &  for  $L_-<x<L - L_+$ \\
$\displaystyle{\frac{3\alpha_+}{R_+}j_+}$ &  for  $L-L_+<x<L$
\end{tabular}
\right.
\end{align}
where the electrolyte flux
\begin{align}
N_\mathrm{e}&(x,t) = -\varepsilon(x)^b D_\mathrm{e}(c_\mathrm{e})\frac{\partial c_\mathrm{e}}{\partial x} \nonumber\\
&-\frac{t^+i(t)}{FA}
\left\{
\begin{tabular}{ll}
$\displaystyle{\frac{x}{L_-}}$ & for  $0<x<L_-$ \\ 
$1$ &  for  $L_-<x<L - L_+$ \\
$\displaystyle{\frac{L-x}{L_+}}$ &  for  $L-L_+<x<L$
\end{tabular}
\right.
\end{align}
with
\begin{align}
    \varepsilon(x)=\left\{
\begin{tabular}{ll}
$\displaystyle{\varepsilon_-}$ & for  $0<x<L_-$ \\ 
$\varepsilon_\mathrm{sep}$ &  for  $L_-<x<L - L_+$ \\
$\displaystyle{\varepsilon_+}$ &  for  $L-L_+<x<L$,
\end{tabular}
\right.
\end{align}
boundary conditions
\begin{align}
\left. \frac{\partial c_\mathrm{e}}{\partial x}\right\vert_{x=0}=0\ \text{ and } \left. \frac{\partial c_\mathrm{e}}{\partial x}\right\vert_{x=L}=0,
\end{align}
and initial condition $c_\mathrm{e}(x,0)=c_{\mathrm{e},0}$.

\paragraph{Terminal voltage} 
\begin{align}
v(t)= &\overbar{v_+} - \overbar{v_-}
+\eta_\mathrm{e}
+ R_0 i(t)
\end{align}
where the electrolyte overpotential
\begin{align}
    \eta_\mathrm{e}(t)&=\frac{2R_\mathrm{g}T(1-t^+)}{F}\overbar{\log\left(\frac{c_{\mathrm{e},+}}{c_{\mathrm{e},-}}\right)}.
\end{align}

\section*{Appendix B: Numerical impedance computation}
\setcounter{equation}{0}
\renewcommand{\theequation}{B.\arabic{equation}}

Software such as \texttt{PyBaMM} simulates a battery by integrating a system of DAEs
\begin{equation}
    \label{eqn:state1}
    \overline{\vec{M}}_\theta\big(\overline{\vec{x}}(t)\big)\frac{\mathrm{d}\overline{\vec{x}}(t)}{\mathrm{d}t} = \overline{\vec{F}}_\theta\big(\overline{\vec{x}}(t), t\big), 
\end{equation}
where $\overline{\vec{M}}_\theta\in \mathbb{R}^{N_\vec{x}\times N_\vec{x}}$ is the mass matrix, which may be singular\footnote{The system \eqref{eqn:state1} is often written in the semi-explicit form $ \frac{\mathrm{d}\vec{y}}{\mathrm{d}t} = \vec{F}(\vec{y},\vec{z},t), 0= \vec{G}(\vec{y},\vec{z},t)$, with $\overline{\vec{x}}=[\vec{y}, \vec{z}]^T$. Here $\vec{y}$ and $\vec{z}$ are the differential and algebraic states, respectively.}, $\overline{\vec{x}}\in \mathbb{R}^{N_\vec{x}\times 1}$ is the vector with discretised states, and $\overline{\vec{F}}_\theta: \mathbb{R}^{N_\vec{x}\times 1}\rightarrow \mathbb{R}^{N_\vec{x}\times 1}$ is a nonlinear vector-valued function. The number of states $N_\vec{x}$ for different models (which depends on the number of discretisation points) are shown in Table~\ref{table:compareMethods}. Both $\overline{\vec{M}}_\theta$ and $\overline{\vec{F}}_\theta$ depend on the model parameters $\theta$. This system of DAEs can be solved using a time-stepping algorithm. Note that some models, such as the single particle model \cite{marquis2019}, result in a system of ordinary differential equations instead of DAEs.

PyBaMM does not always use voltage and current as state variables and therefore we need to rewrite \eqref{eqn:state1} in a form that allows us to compute an impedance. Assuming that the right-hand side $\overline{\vec{F}}_\theta$ depends on $t$ explicitly only through the applied current $i(t)$, the system of DAEs can be rewritten by adding current and voltage to the states,
\begin{equation}
    \label{eqn:state}
    \vec{M}_\theta\big(\vec{x}(t)\big)\frac{\mathrm{d}\vec{x}(t)}{\mathrm{d}t} = \vec{F}_\theta\big(\vec{x}(t)\big) +
   \vec{B}i(t),
\end{equation}
with $\vec{B} = [0,0,..., 1]^\top$ and 
\begin{align}
    \vec{x}^\top(t) = [\overline{x}_1(t),\overline{x}_2(t),\hdots,\overline{x}_{N_\vec{x}}(t),v(t),i(t)]\in \mathbb{R}^{(N_\vec{x}+2)\times 1}.
\end{align}
The last two rows of $\vec{M}_\theta$ are then zeros and the last two rows in $\vec{F}_\theta$ correspond to the algebraic equations. The first of these is the implicit expression for the terminal voltage $v(t)$ in terms of the other state variables $G(\vec{x}(t))=0$ (which depends on the particular model) and the last row imposes that $x_{N_\vec{x}+2}(t) =i(t)$ by setting $F_{N_\vec{x}+2}(\vec{x}(t))=-x_{N_\vec{x}+2}(t)$. In matrix notation we have
\small
\begin{align}
    \begin{bmatrix}
        M_{1,1} & M_{1,2} & \hdots & M_{1,N_\vec{x}+2}\\
        M_{2,1} & M_{2,2} & \hdots & M_{2,N_\vec{x}+2}\\
        \vdots & \vdots &                & \vdots\\
        M_{N_\vec{x},1} & M_{N_\vec{x},2} & \hdots & M_{N_\vec{x},N_\vec{x}+2}\\
        0 & 0 & \hdots & 0\\
        0 & 0 & \hdots & 0
    \end{bmatrix}
    \begin{bmatrix}
        \dot x_1\\
        \dot x_2\\
        \vdots\\
        \dot x_{N_\vec{x}}\\
        \dot v\\
        \dot i
    \end{bmatrix}=\begin{bmatrix}
        F_1(\vec{x})\\
        F_2(\vec{x})\\
        \vdots\\
        F_{N_\vec{x}}(\vec{x})\\
        G(\vec{x})\\
       -i
    \end{bmatrix}+
    \begin{bmatrix}
        0\\
        0\\
        \vdots\\
        0\\
       0\\
       1
    \end{bmatrix}i.
\end{align}
\normalsize
To calculate the impedance from the model, we need to linearise \eqref{eqn:state} and transform the expression into the frequency-domain.

\subsection*{Linearisation}
To linearise the system of equations \eqref{eqn:state}, we Taylor series expand $\vec{F}_\theta(\vec{x}(t))$ around an operating point $\vec{x}_m$ (depending on SOC, temperature, etc.)
\begin{align}
    \vec{F}_\theta(\vec{x}(t))=\vec{F}_\theta(\vec{x}_m)+ \underbrace{\frac{\partial \vec{F}_\theta(\vec{x})}{\partial \vec{x}}\bigg\rvert_{\vec{x}=\vec{x}_m}}_{\vec{J}_{\theta,m}}(\underbrace{\vec{x}(t)-\vec{x}_m}_{\tilde{\vec{x}}(t)})+\text{h.o.t.,}
    \label{eqn:TaylorSeries}
\end{align}
with $\vec{J}_{\theta,m}\in \mathbb{R}^{(N_\vec{x}+2)\times (N_\vec{x}+2)}$ the Jacobian of $\vec{F}_\theta(\vec{x})$, depending on the operating point $\vec{x}_m$, and h.o.t. standing for higher order terms. Note that while the vector-valued function $\vec{F}_\theta$ is dependent on all model parameters $\theta$, the Jacobian $\vec{J}_{\theta,m}$ may not be (that is, model information may be lost during linearisation). The mass matrix can similarly be expanded using a Taylor series
\begin{align}
    \vec{M}_\theta(\vec{x}(t))=\underbrace{\vec{M}_{\theta}(\vec{x}_m)}_{M_{\theta,m}}+ \frac{\partial \vec{M}_\theta(\vec{x})}{\partial \vec{x}}\bigg\rvert_{\vec{x}=\vec{x}_m}(\underbrace{\vec{x}(t)-\vec{x}_m}_{\tilde{\vec{x}}(t)})+\text{h.o.t.}
    \label{eqn:TaylorSeriesM}
\end{align}
Note that the Jacobian of the mass matrix $\vec{M}_\theta(\vec{x})$ is a tensor. Plugging \eqref{eqn:TaylorSeries} and \eqref{eqn:TaylorSeriesM} in \eqref{eqn:state} and neglecting the quadratic and higher order terms in $\tilde{\vec{x}}(t)$ to linearise the system\footnote{Note that this approximation is valid for small current perturbations, which is the requirement for EIS.}, we obtain
\begin{align}
\vec{M}_{\theta,m}&\frac{\mathrm{d}}{\mathrm{d}t}\left(\vec{x}_m+\tilde{\vec{x}}(t)\right)+ \frac{\partial \vec{M}_\theta(\vec{x})}{\partial \vec{x}}\bigg\rvert_{\vec{x}=\vec{x}_m}\tilde{\vec{x}}(t)\frac{\mathrm{d}\vec{x}_m}{\mathrm{d}t}\nonumber \\ 
&=\vec{F}_\theta(\vec{x}_m)+\vec{B}i_m+\vec{J}_{\theta,m}\tilde{\vec{x}}(t) + \vec{B}\tilde i(t),
\label{eq:DAELinearisedOperando}
\end{align}
with $\tilde i(t)=i(t)-i_m$ and $i_m$ the $(N_\vec{x}+2)$-th entry of $\vec{x}_m$. The operating point $\vec{x}_m$ should also satisfy \eqref{eqn:state}, which simplifies \eqref{eq:DAELinearisedOperando} to 
\begin{align}
    \vec{M}_{\theta,m}&\frac{\mathrm{d}\tilde{\vec{x}}(t)}{\mathrm{d}t}+ \frac{\partial \vec{M}_\theta(\vec{x})}{\partial \vec{x}}\bigg\rvert_{\vec{x}=\vec{x}_m}\tilde{\vec{x}}(t)\frac{\mathrm{d} \vec{x}_m}{\mathrm{d}t}=\vec{J}_{\theta,m}\tilde{\vec{x}}(t) + \vec{B}\tilde i(t).
\end{align}
The operating point $\vec{x}_m$ does not necessarily have to be a steady-state, for instance taking $i_m\neq0$ makes it possible to compute the impedance in \textit{operando} conditions \cite{hallemans2022operando,zhu2022operando}, although for this paper we only consider impedance in stationary conditions ($\frac{\mathrm{d}\vec{x}_m}{\mathrm{d}t} = 0$ and $i_m=0$), hence, 
\begin{align}
   \vec{M}_{\theta,m}\frac{\mathrm{d}\tilde{\vec{x}}(t)}{\mathrm{d}t} = \vec{J}_{\theta,m}\tilde{\vec{x}}(t) +
  \vec{B} i(t).
  \label{eqn:linearised}
\end{align}
We have now obtained a linear system of DAEs which is valid for small perturbations at a fixed operating point $\vec{x}_m$ of the battery.

\subsection*{Frequency-domain transformation}
 To transform the linear set of DAEs into the frequency-domain, we take the Fourier transform of \eqref{eqn:linearised} and obtain
 \begin{align}
    \label{eqn:matrixeq}
    j\omega \vec{M}_{\theta,m}\tilde{\vec{X}}(\omega) = \vec{J}_{\theta,m}\tilde{\vec{X}}(\omega)+\vec{B} I(\omega),
\end{align}
where $\tilde{\vec{X}}(\omega)=\mathcal{F}\big\{\tilde{\vec{x}}(t)\big\}$ and using the property of Fourier transforms that
\begin{align}
    \mathcal{F}\Bigg\{\frac{\mathrm{d}\tilde{\vec{x}}(t)}{\mathrm{d}t}\Bigg\}=j\omega \tilde{\vec{X}}(\omega).
\end{align}
Hence, we can write the solution of \eqref{eqn:matrixeq} as  
\begin{align}
    \tilde{\vec{X}}(\omega)=\underbrace{(j \omega \vec{M}_{\theta,m} -\vec{J}_{\theta,m})^{-1}\vec{B}}_{\vec{K}_{\theta,m}(\omega)\in \mathbb{C}^{(N_\vec{x}+2)\times 1}}I(\omega).
    \label{eq:inverseMatrix}
\end{align}

Finally, we find a numerical expression for the impedance scalar at the operating point $\vec{x}_m$ and angular frequency $\omega$ by selecting the $(N_\vec{x}+1)$-th entry of the vector $\vec{K}_{\theta,m}(\omega)$,
\begin{align}
    Z_m(\omega,\theta)=\frac{V(\omega)}{I(\omega)}=\vec{K}_{\theta,m}(\omega)_{[N_\vec{x}+1]}.
    \label{eq:NumericalImpedance}
\end{align}
Finding the vector $\vec{K}_{\theta,m}(\omega)$ involves the inverse matrix operation in \eqref{eq:inverseMatrix} at the selected set of angular frequencies $\omega_k$ so that $Z_m(\omega_k,\theta)$ can be used for the model parameter estimation \eqref{eq:costfunctionImpedance}. The simplest approach for finding this vector is to do a direct solve of \eqref{eqn:matrixeq} (e.g.\ with LU decomposition) which appears to be more computationally efficient than iterative methods (e.g.\ BicgSTAB \cite{vorst1992}) for typical battery models.


\section*{Appendix C: SPMe with grouped parameters}
\setcounter{equation}{0}
\renewcommand{\theequation}{C.\arabic{equation}}

We now write the dimensional SPMe of Appendix~A with grouped parameters. We do this by scaling some of the variables with parameters as per Table~\ref{tab:ScalingVariables}. We decide only to retain the dimensions of time, current, and voltage. The SPMe can be reformulated as follows with the grouped parameters listed in Table~\ref{tab:groupedParamsSPM}.
\begin{table}[]
\centering
\begin{tabular}{ll} \hline 
    Scaled variables & Unit\\ \hline
    $r_\pm=r^*_\pm/R_\pm$ &  dimensionless \\
    $x=x^*/L$ & dimensionless\\
    $c_\pm=c^*_\pm/c_{\pm,\mathrm{max}}$ & dimensionless \\
    $c_\mathrm{e}=c_\mathrm{e}^*/c_{\mathrm{e},0}$ & dimensionless \\$N_\mathrm{e}=N_\mathrm{e}^*/(c_{\mathrm{e},0}L)$ & 1/s \\
    $j_\pm=j^*_\pm/(R_\pm c_{\pm,\mathrm{max}})$ & 1/s\\
    $i_{0,\pm}=i^*_{0,\pm}/(F R_\pm c_{\pm,\mathrm{max}})$ & 1/s\\
 \hline
 \end{tabular}
 \caption{Scaling of the variables of the full model. The variables of the full models are denoted with an asterisk in superscript and the new variables without. Other variables are kept the same as in the full dimensional model.}
 \label{tab:ScalingVariables}
 \end{table}
 
 \paragraph{Diffusion in the spherical solid particles in each of the electrodes}
\begin{align}
\frac{\partial c_{\pm}}{\partial t} = \frac{1}{r^2}\frac{\partial}{\partial r}\left(\frac{r^2}{\tau^\mathrm{d}_{\pm}}\frac{\partial c_{\pm}}{\partial r} \right)\qquad 0\leq r\leq 1
\end{align}
with boundary conditions
\begin{align}
\left.\frac{\partial c_{\pm}}{\partial r}\right\vert_{r=0}= 0, \qquad \left. -\frac{1}{\tau^\mathrm{d}_{\pm}}\frac{\partial c_{\pm}}{\partial r}\right\vert_{r=1}= \overbar{j_{\pm}},
\end{align}
and initial conditions 
\begin{align}
    c_{\pm}(r,0)=c_\pm^{0\%}+\frac{\text{SOC}_0}{100}\left(c_\pm^{100\%}-c_\pm^{0\%}\right).
\end{align}

\paragraph{Intercalation reaction at the particle surface}
\begin{align}
j_\pm(x,t) = 2i_{0,\pm}\sinh\Bigg(\frac{F\eta_\pm}{2R_\mathrm{g}T}\Bigg)
\label{eq:overpotentialGrouped}
\end{align}
with
\begin{align}
i_{0,\pm} &= \frac{1}{\tau^\mathrm{ct}_\pm}\sqrt{c_\pm|_{r=1}}\sqrt{c_\mathrm{e,\pm}(1-c_\pm|_{r=1})}, \\
\eta_\pm(x,t) &= v_{\pm} -U_{\pm}(c_{\pm}|_{r=1}), \\
v_{\pm}(x,t) &= \overbar{v_{\pm}}+ \frac{2R_\mathrm{g}T}{F}(1 - t^+)\left(\overbar{\log(c_{\mathrm{e},\pm})} -\log(c_{\mathrm{e},\pm}) \right).
\label{eq:vpmGrouped}
\end{align}

\paragraph{Transfer process at the particle surface with double-layer} 
\begin{align}
C_\pm\frac{\mathrm{d} \overbar{v_{\pm}}}{\mathrm{d}t} = \pm i(t) - 3Q_\pm^\mathrm{th}\overbar{j_{\pm}}
\label{eq:doubleLayerGrouped}
\end{align}
with initial condition $\overbar{v_{\pm}}(0)=U_{\pm}(c_{\pm, 0})$.

\paragraph{Diffusion in the electrolyte} 
\begin{align}
\zeta(x)\frac{\partial c_\mathrm{e}}{\partial t} = &-\frac{\partial N_\mathrm{e}}{\partial x} \nonumber \\
&+\frac{3}{Q^\mathrm{e}}\left\{
\begin{tabular}{ll}
$\displaystyle{\frac{Q^\mathrm{th}_-j_-}{\ell_-}}$ &  for  $0<x<\ell_-$ \\ 
$0$ &  for  $\ell_-<x<1 - \ell_+$ \\
$\displaystyle{\frac{Q^\mathrm{th}_+j_+}{\ell_+}}$ & for  $1-\ell_+<x<1$
\end{tabular}
\right.
\end{align}
where $Q^\mathrm{th}_\pm$ are given by \eqref{eq:theoreticalElectrodeCapacitySto} and the electrolyte flux
\begin{align}
    N_\mathrm{e} = &-\frac{1}{\tau_\mathrm{e}(x)}\frac{\partial c_\mathrm{e}}{\partial x} \nonumber \\
    &-\frac{t^+i(t)}{Q^\mathrm{e}}
    \left\{
\begin{tabular}{ll}
$\displaystyle{\frac{x}{\ell_-}}$ &  for  $0<x<\ell_-$ \\ 
$1$ &  for  $\ell_-<x<1 - \ell_+$ \\
$\displaystyle{\frac{1-x}{\ell_+}}$ & for  $1-\ell_+<x<1$
\end{tabular}
\right.
\end{align}
with 
\begin{align}
    \zeta(x)=&\left\{
\begin{tabular}{ll}
$\zeta_-$ &  for  $0<x<\ell_-$ \\ 
$1$ &  for  $\ell_-<x<1 - \ell_+$ \\
$\zeta_+$ & for  $1-\ell_+<x<1$,
\end{tabular}
\right.
\\
    \tau_\mathrm{e}(x)=&\left\{
\begin{tabular}{ll}
$\tau^\mathrm{e}_-$ &  for  $0<x<\ell_-$ \\ 
$\tau^\mathrm{e}_\mathrm{sep}$ &  for  $\ell_-<x<1 - \ell_+$ \\
$\tau^\mathrm{e}_+$ & for  $1-\ell_+<x<1$,
\end{tabular}
\right.
\end{align}
boundary conditions
\begin{align}
\left. \frac{\partial c_\mathrm{e}}{\partial x}\right\vert_{x=0}=0, \qquad \left. \frac{\partial c_\mathrm{e}}{\partial x}\right\vert_{x=1}=0,
\end{align}
and initial condition $c_\mathrm{e}(x,0)=1$.

\paragraph{Terminal voltage}
\begin{align}
v(t)= &\overbar{v_+} - \overbar{v_-} +\eta_\mathrm{e} + R_0 i(t)
\end{align}
where the electrolyte overpotential
\begin{align}
    \eta_\mathrm{e}(t)=&\frac{2R_\mathrm{g}T}{F} (1-t^+)\overbar{\log\Bigg(\frac{c_{\mathrm{e},+}}{c_{\mathrm{e},-}}}\Bigg).
\end{align}

\section*{Appendix D: Charge transfer resistance}
\setcounter{equation}{0}
\renewcommand{\theequation}{D.\arabic{equation}}
The charge transfer resistance at the operating point $x_m$ is the derivative of the voltage contribution from the spatially averaged overpotential $\overbar{\eta_\pm}$,
with respect to the current, evaluated at the operating point $x_m$,
\begin{align}
R_{\mathrm{ct},m,\pm}
&=\pm \frac{\mathrm{d}\overbar{\eta_\pm}}{\mathrm{d}i}\bigg\rvert_{x_m}.
\end{align}
In general, the average overpotential does not correspond to the average molar flux $\overbar{j_\pm}$ but, from \eqref{eq:overpotentialGrouped} and \eqref{eq:vpmGrouped},
\begin{align}
j_\pm = 2i_{0,\pm}\sinh\Bigg(\frac{F\overbar{\eta_\pm}}{2R_\mathrm{g}T} + (1 - t^+)\left(\overbar{\log(c_{\mathrm{e},\pm})}-\log(c_{\mathrm{e},\pm})\right)\Bigg).
\end{align}
For small currents, we linearise as
\begin{align}
j_\pm \approx 2i_{0,\pm}\Bigg(\frac{F\overbar{\eta_\pm}}{2R_\mathrm{g}T} + (1 - t^+)\left(\overbar{\log(c_{\mathrm{e},\pm})}-\log(c_{\mathrm{e},\pm})\right)\Bigg).
\end{align}
In practice, we find that the contribution from the spatial variation of the electrolyte stoichiometry is much smaller than the overpotential, therefore,
\begin{align}
\overbar{j_\pm} \approx 2\overbar{i_{0,\pm}}\Bigg(\frac{F\overbar{\eta_\pm}}{2R_\mathrm{g}T}\Bigg)
= \frac{F\overbar{\eta_\pm}}{R_\mathrm{g}T}\overbar{i_{0,\pm}}.
\end{align}
Noting that $x_m$ is a state at which $i=0$,
\begin{align}
R_{\mathrm{ct},m,\pm}
&\approx \pm \frac{R_\mathrm{g}T}{F\overbar{i_{0,\pm}}}\frac{\mathrm{d}\overbar{j_\pm}}{\mathrm{d}i} \bigg\rvert_{x_m}.
\end{align}
From  \eqref{eq:doubleLayerGrouped} (neglecting the double-layer capacitance) this yields the conventional result that 
\begin{align}
R_{\mathrm{ct},m,\pm}
&\approx \frac{R_{\mathrm{ct},\pm}^\mathrm{typ}}{2\sqrt{c_{m,\pm}(1-c_{m,\pm})}}\frac{1}{\overbar{\sqrt{c_{\mathrm{e},\pm}}}\big\rvert_{x_m}}.
\label{eq:Rct_app}
\end{align}

\section*{Acknowledgements}
This research was supported by the Faraday Institution Nextrode (FIRG066) and Multiscale Modelling (MSM) (FIRG059) projects, as well as the EU IntelLiGent (101069765) project through the UKRI Horizon Europe Guarantee (10038031). OCP data of the LG~M50LT cell was provided by About:Energy Ltd (London, UK, \url{https://www.aboutenergy.io/}). For the purpose of Open Access, the authors have applied a CC BY public copyright licence to any Author Accepted Manuscript (AAM) version arising from this submission. 

\section*{Data and code availability}
\noindent \texttt{PyBAMM-EIS}:\\
\url{https://github.com/pybamm-team/pybamm-eis}\\\\
\texttt{PyBOP}: \\
\url{https://github.com/pybop-team/PyBOP}
\section*{Competing Interests}
D.A.H.\ is co-founder of Brill Power Ltd. 
The other authors have no competing interests to declare.

\bibliographystyle{elsarticle-num}
\bibliography{paper_NH}

\end{document}